\documentclass[letter,11pt,titlepage]{report}
\usepackage{amssymb}
\usepackage{amsmath}
\usepackage{epsfig}
\usepackage{verbatim}

\title{Detection of a Hypercharge Axion in ATLAS\\
	{\large
	a Monte-Carlo Simulation of\\
	a Pseudo-Scalar Particle (Hypercharge Axion)\\
	with Electroweak Interactions\\
	for the ATLAS Detector\\
	in the Large Hadron Collider at CERN\\}
}
\author{Erik~Elfgren\\
	elf@ludd.luth.se}
\date{December, 2000\\
	Division of Physics\\
	Lule\aa University of Technology\\
	Lule\aa, SE-971 87, Sweden\\
	http://www.luth.se/depts/mt/fy/}

\newcommand{\gloss}[1]{{\sl #1}}

\begin{document}
        \maketitle

\begin{abstract}
This Master of Science thesis treats the hypercharge axion, which is a hypothetical pseudo-scalar particle with
         electroweak interactions.

First, the theoretical context and the motivations for this study are discussed. In short,
         the hypercharge axion is introduced to explain the dominance of matter over antimatter in the universe and the
         existence of large-scale magnetic fields.

Second, the phenomenological properties are analyzed and the
         distinguishing marks are underlined. These are basically the products of photons and $Z^0$s with 
	 high transverse momenta and invariant mass equal to that of the axion.

Third, the simulation is carried out with two photons producing the
         axion which decays into $Z^0$s and/or photons. The event simulation is run through the simulator ATLFAST of
         ATLAS (A Toroidal Large Hadron Collider ApparatuS) at CERN.

Finally, the characteristics of the axion decay
         are analyzed and the criteria for detection are presented. A study of the background is also included. The result is
         that for certain values of the axion mass and the mass scale (both in the order of a TeV), the hypercharge axion
         could be detected in ATLAS.
\end{abstract}

\chapter*{Preface}
This is a Master of Science thesis at the Lule\aa~University of Technology, Sweden.
The research has been done at Universit\'e de Montr\'eal, Canada, under the
supervision of Professor Georges Azuelos. My thesis receiver in Sweden has been
Professor Sverker Fredriksson.

The thesis is divided into three chapters. The first one treats the
motivations for the hypercharge axion and some theoretical background.
The second covers the preparations and simulations as well as some
predictions from the theory. The third chapter contains the results
of the simulations and some conclusions.

Notations and units: In the calculations we use units where Planck's
constant, the speed of light and Boltzmann's constant are all
equal to unity.
A bar over a particle name signifies
the antiparticle.
The word axion is interchangeable with hypercharge axion.
Footnotes are used to explain further, but are not necessary for the
basic comprehension.
If a word is marked in \gloss{slanted} it is either supposed to
be known, or is {\it curiosum}.
These words are explained in the Glossary
in Appendix \ref{app:Gloss}. All symbols, abbreviations and constants used in the thesis are
listed and briefly explained in Appendix \ref{App:Symb}.

Finally I would like to express my deep gratitude to all the people that
have helped on the different subjects of this thesis. First of all to my
supervisor Georges Azuelos, Universit\'e de Montr\'eal, for his eternal
patience with all my questions as well as numerous suggestions and good
discussions. I would also like to thank Gilles Coutures, Universit\'e de
Qu\'ebec de Montr\'eal, who has been
a great help and who made the theoretical work leading to this thesis.
I thank my professor in Sweden, Sverker Fredriksson, Lule\aa~University
of Technology, for his support and
all our previous discussions that have helped me in this work.
As for the motivations for the hypercharge axion, I would like to thank
Roger MacKenzie, Robert Brandenberger, James Cline, and Salman Habib for interesting
discussions on cosmology, sphalerons and electroweak baryogenesis.

Montr\'eal in December 2000,
Erik Elfgren.
        \tableofcontents

\chapter*{Introduction}
Ever since the discovery of antimatter it has been a
mystery that almost all of our experiments in particle
physics are symmetric in matter and antimatter,
but yet the universe seems to be constituted entirely
of matter. Some general conditions were outlined by
Sakharov in 1967, but the problem itself remains unsolved.

The first possible explanations were based on the grand unified theories
(GUTs) in which the asymmetry was generated very close
to big bang. These theories allow baryon-to-lepton decay,
which could generate an asymmetry. One drawback is that
the theories cannot be tested without tremendous amounts
of energy ($\gtrsim 10^{16}$ GeV), far beyond our reach.

Later on, theories evolved that could explain the generation
of the asymmetry without GUT theories. Most of them
suppose that the asymmetry was created around the electroweak
phase transition at $T\sim 100$ GeV when the electroweak
symmetry was broken. Supersymmetric theories can offer
possible explanations under certain conditions, but
the standard model itself seems incapable to produce
the required asymmetry.
 
A theory proposed by Brustein and Oaknin is particularly
appealing for several reasons. It suggests simply that
the introduction of a scalar field could create the
required asymmetry through coupling to the hypercharge.
As a bonus, the theory could also explain the existence
of large magnetic fields in cosmic plasmas.
The pseudoscalar was named hypercharge axion from its couplings
to hypercharge. 

In summary, the hypercharge axion can possibly explain
the dominance of matter over antimatter in the universe.
In other words, why we exist!

Finally, the hypercharge axion is expected to have a mass in
the TeV range, making it possible to detect in the ATLAS
detector at the large hadron collider (LHC) which is now
constructed at CERN, Geneva, and will be operational by 2005.
This is the subject of this masters thesis.

Some references on particle physics are Nash \cite{Nash},
Peskin and Schroeder \cite{PeskSch},
and the CERN photo-gallery: http://press.web.cern.ch/Press/Photos/.

\chapter{Background and Motivations
}
\label{ch:BaMo}
This chapter treats the background to my research, why it may be
of interest to study the axion and what has been done so far.
Briefly, the axion could be the reason for the mysterious domination
of matter over antimatter in the universe. Before introducing the axion
some theory is presented to situate it in its context and 
explain how the axion can be the reason behind the domination of matter.

This chapter is divided into three major parts. The first one addresses
some general questions, such as how we can be sure that the universe really
has more matter than antimatter, and some very general conditions that must be fulfilled
for the asymmetry to exist. These conditions are called the Sakharov criteria
and they must be fulfilled for statistical reasons.

The second part of this chapter is devoted to \gloss{baryogenesis}, i.e., the
process by which the asymmetry is created. The subject is very complicated, so
I will content myself with a rather qualitative description of the different
phenomena. Among these are the \gloss{electroweak phase transition}, which took place
about $10^{-10}$ seconds after the big bang; the role of
hypermagnetic fields in the phase transition; sphalerons which
counter asymmetries and the Chern-Simons number, which is directly proportional
to the baryon asymmetry.

The third part treats the axion itself, with the properties proposed
by Brustein and Oaknin \cite{BruOakColl}.
A description of how the axion could amplify the hypermagnetic fields
is included as well as the couplings and branching ratios.
A discussion of where the hypercharge axion could appear is also provided. 
Its properties are compared to those of particles in the most popular extensions
of the standard model: the minimal supersymmetric model and superstring models.


\section{Matter-Antimatter Asymmetry}
\subsection{How do we know that the Universe is Asymmetric?}
While our theories in particle physics are very nearly symmetric in matter-
antimatter (almost all relevant experiments in particle physics show this behavior),
the universe itself is far from being so.
We have several reasons to believe that the universe is mainly constituted
of matter. The most obvious reason is that we have explored our solar
system without annihilating, which would not have been the case if there had been
antimatter in significant abundance in our solar system.

	To conclude that the universe at large
is made of matter is not as obvious. What if another part of our galaxy,
or another galaxy nearby were constituted of antimatter? This possibility is
improbable since cosmic radiation in that case should contain antimatter.
Cosmic radiation does contain some antiprotons, but that fraction ($10^{-4}$)
is well explained by the reaction $p+p\rightarrow 3p + \bar{p}$,
which has nothing to do with the eventual existence of larger chunks of antimatter.
This is evidence that there is no large antimatter cluster
in our relative proximity.

As for more distant galaxies, we should see gamma-ray emissions from the 
interface  between matter and antimatter galaxies.
The gamma rays
should have an energy up to that of a particle-antiparticle pair. These gamma
rays should produce a detectable background, which is not observed. Causality
is not an argument against this either, as we can see (almost) all parts
of the universe, only at different times. The separation should
have occurred rather early for any significant amount of antimatter to remain,
so that we should observe gamma rays from these parts too. Thus, there
is a negligible amount of antimatter on the scale of clusters.
For the interested reader, references like Steigman \cite{Steigman},
Stecker \cite{Stecker} and Cohen et al. \cite{Cohen} are recommended.




\subsection{How did the Universe become Asymmetric?}
The first possiblility is that the universe started asymmetric.
This is not a very interesting possibility because of its sterility.
It does not suggest any new physics, nor any "real" explanations or
predictions.
Besides, it is in contradiction with inflation theory, which says
that any initial abundances are diluted.

The second possibility is that the universe started symmetric
(or at least became symmetric at a rather early stage, e.g.~during
inflation), but
the symmetry was broken through some mechanism. For the baryon asymmetry
this mechanism is called baryogenesis. Note that even though baryon
number and lepton number separately are violated, the difference $B-L$ is
thought to be constant.

The imbalance in the amount of matter and antimatter can basically be
measured as
the imbalance in the amount of baryons and antibaryons, since baryons
(neutrons and protons) are the heaviest of the basic constituents of
normal matter. Also, we cannot measure the lepton asymmetry because we 
know neither how many neutrinos there are in the universe, nor their mass.
The asymmetry is often characterized by the baryon-to-entropy (\cite{PDG}, page 133) ratio:
\begin{equation}
	\eta = \frac{n_B}{s} = \frac{n_b - n_{\bar b}}{s} \approx 10^{-10},
\end{equation}
which can be seen as the the ratio between the number of baryons and
the number of photons (which come from the annihilation of matter and
antimatter).

In 1967, Sakharov \cite{Sakharov} stated that there are three criteria
that must be satisfied to explain the present matter-antimatter asymmetry:
\begin{enumerate}
\item	{\bf Non-conservation of baryons}\\
	If the baryon number is conserved in all reactions then the observed 
	asymmetry can only reflect asymmetric initial conditions.
	Hence, there must have existed some process that violated
	the baryon number.
	Grand unified theories (GUTs) have leptons and quarks in the
	same representation, and therefore allow decay of baryons to leptons.
	A process that has been searched for is proton decay to positron,
 	$p \rightarrow e^+\nu_e$, but no evidence
	for such processes has been found so far.
	However, baryon number violation can also be obtained from anomalies,
	which can occur without any extension from the standard model,
	see below.
\item	{\bf Charge ($C$) and Charge-Parity ($CP$) violation}\\
	$CP$ \gloss{symmetry} implies that if there are two identical \gloss{decay channels}, 
	except that one involves matter and the other involves antimatter,
	then both happen with equal probability, e.g.,~
	$P(e^-p^+\rightarrow n\nu) = P(e^+\bar p\rightarrow \bar n\bar\nu)$.
	In other words, if
	$CP$ symmetry is violated and if you could contact someone in another 
	part of the universe you could instruct him how to make
	an experiment to tell whether he is made of matter
	or antimatter.
	Without a preference for matter or antimatter, baryon
	non-conservation would work as much one way as the other,
	and matter and antimatter would have annihilated.
\item	{\bf Thermodynamic nonequilibrium}\\
	The universe must have evolved from a state of thermodynamic
	equilibrium to thermodynamic nonequilibrium.
	This is necessary because in thermodynamic equilibrium the net number
	density of baryons is equal to that of antibaryons. This can be illustrated
	as follows.
	In thermodynamic equilibrium the number density $n(E)$ of a particle is given by
	\begin{equation}
	\label{eq:density}
		n(E) \propto e^{-k_BH/T},
	\end{equation}
	where $k_B$ is Boltzmann constant, $T$ is the temperature and
	$H$ is the Hamiltonian of the system. 
	Now, due to the Charge-Parity-Time ($CPT$) theorem \cite{Kabir} the masses of
	particles and antiparticles are the same\footnote{
		The validity of the $CPT$ theorem is very fundamental, but is
		nevertheless questioned, see for example Eades and Hartman \cite{Eades}.
	},
	which means that the Hamiltonians are the same. 
	In other words, equation (\ref{eq:density}) implies that in thermodynamic
	equilibrium, the number (density) of baryons is the same as
	the number (density) of antibaryons.
	You can also explain this condition with the fact that a difference
	in the number of particles and antiparticles decreases the randomness
	of the system, which is prohibited in thermodynamic equilibrium, as the
	second law of thermodynamics states that the entropy of a system must
	increase, or at best stay the same.
	
	One way of bringing the universe out of thermodynamic equilibrium is by a
	phase transition. However, the thermodynamic nonequilibrium needs to be
	very strong to allow the current asymmetry between matter and antimatter.
	A second-order phase transition is charaterized by continuously varying
	parameters, giving a weak thermodynamic nonequilibrium. A first-order
	phase transition has discontinuously varying parameters. To create the
	currently observed matter-antimatter asymmetry, a {\it strong}
	first-order phase transition is needed.


\end{enumerate}

In summary, the entire production of the asymmetry should go trough the following
steps (assuming the simplest procedure):
\begin{enumerate}
	\item	The universe started symmetric in matter and antimatter.
	\item	Processes violating baryon number, $C$ and $CP$ symmetry
		came into action.
	\item	The universe was (temporarily) brought out of thermodynamic equilibrium,
		allowing for an asymmetry between matter and antimatter to develop.
	\item	Very shortly after (or during) the return to thermodynamic
		equilibrium the baryon number violating processes stopped (otherwise the 
		return to thermodynamic equilibrium would once again have brought the universe to
		symmetry between matter and antimatter).
	\item	The remaining matter and antimatter annihilated (into photons)
		until only the surplus matter remained.
\end{enumerate}

\section{Electroweak Model of Baryogenesis}
In the minimal standard model, two of the Sakharov criteria are problematic.
First, the high mass of the Higgs boson seems to preclude a first-order
phase transition \cite{EWBaGe}, which would have been a natural way to
bring the universe out of thermodynamic equilibrium. Second, the CP violation
is too weak to produce the observed baryon-to-photon ratio $10^{-10}$.
This means that extensions of the standard model need to be considered,
but before doing that an explanation of how the asymmetry can be created
is appropriate.


Baryogenesis should take place sometimes very early in the history of the universe, in the
first fractions of a second, before the formation of compounds of quarks like baryons
and mesons (note that "free" quarks also have a baryon number).

As the baryogenesis must take place in thermodynamic nonequilibrium, there are
two basic possibilities to investigate. First of all, the pure expansion of
the universe provokes a thermodynamic nonequilibrium. However, at the time around
the electroweak phase transition this has been estimated to be by far too weak to stop the symmetry from
being restored. Second, the thermodynamic nonequilibrium can be caused by a phase
transition, like the electroweak phase transition $10^{-10}$ seconds after the Big Bang.
Thus the thermodynamic nonequilibrium must either be caused by a
phase transition, or occur at a very early stage at the scale of the grand unified theories (GUTs). 
In this study, the electroweak phase transition will be treated in further detail.

Some aspects of the GUT thermodynamic nonequilibrium can be found in
Kolb and Wolfram \cite{Kolb}.

\subsection{Analogy with a Pendulum}
\label{sec:Pendulum}
For several of the concepts introduced in the following sections, an analogy with a
mechanical pendulum can be of great help. The Lagrangian of the pendulum is:
\begin{equation}
	L = \textstyle{\frac 12}ml^2\dot\theta^2 - mgl(1-\cos \theta),
\end{equation}
where $m$ is the mass and $l$ is the length of the pendulum ($\theta$ is the angle
of inclination of the pendulum with respect to the vertical and $g\approx
9.8$ m/s$^2$ is the standard acceleration of gravity).

The system is periodic and the periodicity can be labeled by an integer $n$:
\begin{equation}
	\theta_n = 2n\pi.
\end{equation}
Below is a list of concepts that have anlogues for the electroweak phase
transition:

\begin{description}
	\item[Winding number]
		The number of rotations, $n$.
	\item[Vacuum]
		The solution where the pendulum is at rest ($\theta = 0$).
	\item[Sphaleron]
		The solution where the pendulum is at its highest point ($\theta~=~\pi$).
		Sphaleron is also often used as a label of a rotation, passing
		the sphaleron point, going to an adjacent winding number $n$.
		The energy of the sphaleron point is $mgl(1-\cos \pi) = 2mgl$.
	\item[Instanton]
		The tunneling of the pendulum from one winding number to the adjacent one.
		(This is a quantum mechanical effect and does not really make sense
		in this mechanical context, but it is included for completeness.)
	\item[Thermal energy]
		A set of pendula having a randomly distributed energy, but with the
		mean energy equal to the thermal energy.
\end{description}
If the thermal energy becomes comparable to the sphaleron energy, it becomes
possible for thermal transitions over the energy barrier to occur. The 
rate of these transitions can be calculated to be
\begin{equation}
	\Gamma(T) \propto e^{-E_{sph}/T},
\end{equation}
so that when $T \sim E_{sph}$ the transitions between different vacua (different winding numbers)
happen unhindered.

Another analogy between the pendulum and quantum field theory is that for low
$T$, i.e., $\theta\approx 0$, we can approximate
$\cos \theta \approx 1 - \theta^2/2$, which is the same thing as perturbation theory, which
also is a kind of Taylor expansion around a minimum in the potential. However,
in both cases information about the periodic structure of the system is lost.

\subsection{Thermodynamic Nonequilibrium}
This is the era when the Higgs first acquired a vacuum expectation
value (vev), so the $Z^0$ and $W^\pm$ acquired masses. The temperature of the universe
was at this stage of the order of 100 GeV.

As mentioned earlier, the phase transition has to be of first-order to induce a
thermodynamic nonequilibrium strong enough. In fact, it is not even sufficient
with a first order phase transition, but a {\it strong} first order phase
transition is required in the simplest model.

The dynamics of the phase transition is obviously very important, and the imagined
procedure is in analogy with that of pressurized boiling water. Close to a critical
temperature $T_c$, ``bubbles'' of the other phase starts to form. These are regions of 
space where the electroweak symmetry is broken. For a start these bubbles die out as
fast as they appear, but then they grow fast enough to form nucleates (called critical bubbles)
and they expand and eventually fill the entire universe. At the bubble walls there
is a significant departure from thermodynamic equilibrium.

\subsection{Nonconservation of Baryon Number}
\begin{figure}
        \centering
        \includegraphics[width=3in]{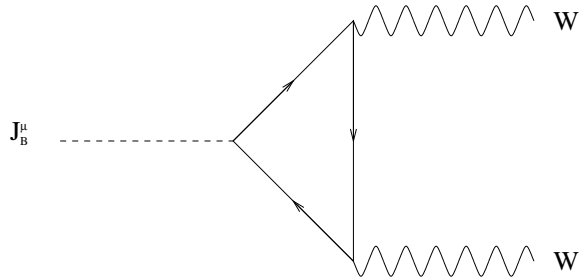}
        \caption{The triangle anomaly contributing to the baryon and lepton number
	currents, and thereby the violation of baryon and lepton number.}
       \label{fig:triangle}
\end{figure}

Classically\footnote{
	Classically means with only perturbative calculations.
} the Lagrangian conserves both the baryon number and the lepton number.
However, there are triangle anomalies (see figure \ref{fig:triangle}),
which add radiative corrections to the Lagrangian. The theory can be renormalized
with conservation of either the vector {\it or} the \gloss{axial current}, but
not both. We know very
well that the vector current is conserved (by conservation of charge), which
means that the axial current is not conserved. Now the integral over time of the divergence of 
the axial current (along with the conservation of the vector current) implies a
change in baryon number.\footnote{
        Compare with classical electron current $\int_t\nabla\cdot \vec J = \Delta Q$, where
        $\vec J$ is the current, $t$ is time and $\Delta Q$ the change in charge.
}

However, the radiative corrections from the anomalies are non-perturbative and at the 
temperature of the universe today, the corrections are extremely small. In the case
of the pendulum above, they correspond to a thermal excitation such that
the pendulum would pass its highest point, the \gloss{sphaleron}, or they correspond to
\gloss{instanton} tunneling.


As in the case of the pendulum above, there is also a \gloss{winding number}
associated with
the Lagrangian, the \gloss{Chern-Simons winding number}, $N_{CS}$.
The relation between the Chern-Simons number and the change in baryon number can be expressed as:
\begin{equation}
	\Delta B = \Delta N_{CS} = n_f[N_{CS}(t_1)-N_{CS}(t_0)],
\end{equation}
where $n_f$ is the number of families and $N_{CS}(t)$ is the Chern-Simons (winding) number at time $t$.
Hence, when the Chern-Simons number is changed by one, nine quarks, (three color states for
each generation) and three leptons are created, giving $\Delta B = 3$ and $\Delta L=3$.
This change in Chern-Simons number is vehicled
by either a \gloss{sphaleron} process or an \gloss{instanton}. Without pretending to
know the details, I give the basic procedure as follows. 

\begin{figure}
        \centering
        \includegraphics[width=5in]{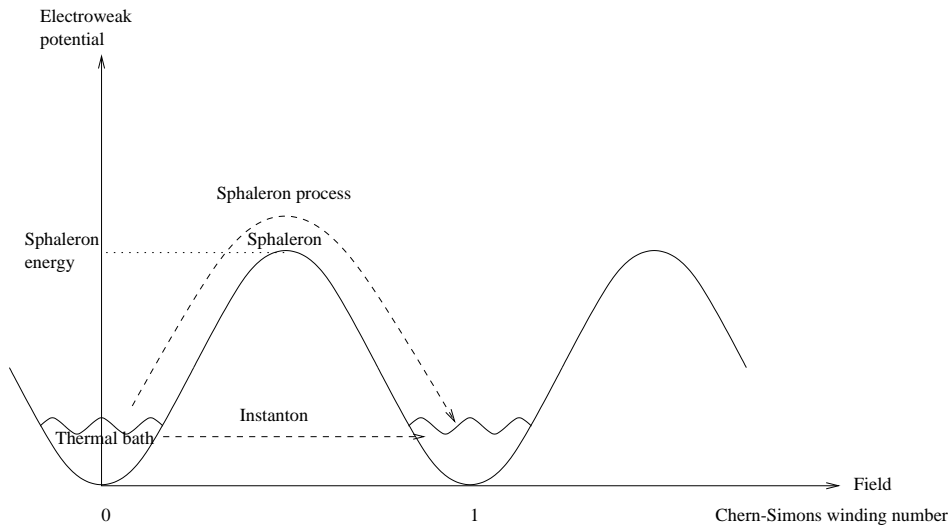}
	\caption{Sketch of electroweak potential with minima,
	Chern-Simons number, sphaleron and instanton.}
	\label{fig:Potential}
\end{figure}

The electroweak potential can be sketched as in figure \ref{fig:Potential}.
There are several different vacua, like in the case of the pendulum, and the different vacua have
different Chern-Simons number (like $n$ in the case of the pendulum).
A transition between different vacua (and therefore different
Chern-Simons number and hence a change in baryon number) can be obtained in two different ways.
First, a tunneling event may occur, which is called an instanton. The probability of tunneling
depends on the thermal energy, i.e., how "thick" the potential barrier between different
Chern-Simons numbers is.
Second, if the (thermal) energy is high enough, it is possible to pass above the potential walls
separating the different Chern-Simons numbers (and different vacua).
The solution at the highest point of the potential
is called a sphaleron, but also a transition that {\it passes} over this point is called a sphaleron
(or a sphaleron process).
These two processes will be outlined in the following sections.\footnote{
	Nonlinear coherent structures such as kinks, solitons, sphalerons, 
	vortices, and instantons arise in the study of a variety of interesting and
	important physical situations such as baryon number violation, properties of materials,
	optical communications, dynamics of phase transitions, and the physics of the early universe.
}

\subsubsection{Sphalerons}
Mathematically, a sphaleron is a special type of solution to a partial differential equation.
It is a saddle-point solution to the equations of motion with a single negative eigenvalue.

It corresponds physically to the solution where the energy of the state is equal to that
of the highest point of the (periodic) potential. For the electroweak potential, the
energy of the spahleron can be calculated \cite{Klinkhammer} to be
\begin{equation}
	E_{sph}(T\approx 0) \approx 10 \textrm{ TeV}.
\end{equation}
The result is only an approximation, but it still gives an idea of the order of
magnitude of the sphaleron energy.

Note that the energy of the sphaleron actually depends on the temperature. The sphaleron
energy decreases with increasing temperature, which means that when the temperature 
approaches the electroweak phase transition temperature $\sim 100$ GeV, the sphaleron
energy is believed to discontinuously go to zero.

The rate per volume of baryon number violating effects can be approximated to be
\begin{equation}
	\Gamma(T) \propto e^{-E_{sph}(T)/T}
\end{equation}
for a certain regime \cite{EWBaGe} (probably) below the electroweak phase transition.
This is the same type of transition rate formula as for the pendulum, and we
expect that near the critical temperature the 
exponential suppression factor will take effect \cite{Moore} (when decreasing the temperature).


As can be seen in the previous formula,
the sphaleron processes will work against the production of a baryon-antibaryon
asymmetry.
In the scenario above, these sphaleron processes can be seen as the reason for the
necessity of a thermodynamic non-equilibrium because the sphalerons vehicle
the change in baryon number.



\subsubsection{Instantons}
Instantons are, like sphalerons, special types of solutions to partial differential equations.
Intuitively, they can be described as an energy configuration in space-time.
Physically, they describe the tunneling between different Chern-Simons numbers/vacua.
The tunneling rate betweeen different Chern-Simons numbers can be calculated to be
\begin{equation}
	\Gamma \propto e^{-S_E},
\end{equation}
where $S_E$ is the Euclidian action obtained from the Minkowski-space action by
a Wick rotation $t \rightarrow -it \equiv \tau$.
The tunneling rate at zero temperature can be calculated \cite{EWBaGe} to be
\begin{equation}
	\Gamma(T=0) \sim 10^{-170}.
\end{equation}
Thus, in our everyday life, instanton processes will play a very insignificant role,
and baryon number is virually conserved.

\subsection{Charge and Charge-Parity Violation}
In the previous section a rather detailed study of the baryon number violation
has been done because it is closely related to the subject of this report,
the hypercharge axion. As for the $CP$ violation, it has nothing to do with
the axion, and only some brief remarks will be made.

Even though $C$ and $CP$ symmetries are known to be broken in the standard
model, this source of $CP$ violation is far too small to produce the observed
asymmetry between matter and antimatter.

In the minimal supersymmetric standard model (MSSM) $CP$ violation can be enhanced
by including a second Higgs doublet or by higher dimension operators,
but this subject is too vast to be treated in this report. For a resum\'e of
these effects, see \cite{EWBaGe}. This enhancement of the $CP$ violation could be strong enough
to produce the present dominance of matter over antimatter.

\section{The Axion}
\label{sec:Axion}
The indroduction of the hypercharge axion has two objectives. First, it
explains the baryon number violation that gives rise to the matter-antimatter
asymmetry in the universe. Second, it explains the existence of large-scale
magnetic fields in diffuse astrophysical plasmas.
The first Sakharov condition is the principal subject of this study of
the axion.
The second Sakharov condition, the presence of $C$ and $CP$ violation is expected to
be explained elsewhere and the third Sakharov condition is supposed to
be fulfilled by a strong first-order electroweak phase transition.\footnote{
	According to \cite{BruOakAmp} long-range uniform magnetic fields
	could strengthen the electroweak phase transition to the point
	that the baryon number asymmetry could survive.
}

The basic idea of the procedure from axion to baryon number violation is the following:
The hypercharge axion couples to the hypermagnetic field, amplifying it to the extent that
it gives sufficient baryon number violation to account for the matter-antimatter
asymmetry. 

This section covers the theoretical background and properties of
the hypercharge axion.

%
%

\subsection{Amplification of Hypermagnetic Fields}
In the universe today, we observe mysterious large-scale magnetic
fields. These could originate from hypermagnetic fields that existed
before the electroweak phase transition. The hypermagnetic fields
are thought to behave basically like the ordinary electromagnetic fields,
obeying Maxwell's equations. With some different compositions
there still exists two parts, one weak (SU(2)$_L$) and one hyperelectromagnetic
(U(1)$_Y$).

The part of the baryon number violating term, which is important for the hypercharge
axion, is expressed as
\begin{equation}
\label{eq:divJB}
        \partial_\mu J_B^\mu \propto
		Y_{\mu\nu}\tilde{Y}^{\mu\nu} \propto \vec H_Y \cdot \vec E_Y
\end{equation}
because the hypercharge axion couples to the hyper(electro)magnetic fields,
$H_Y$ and $E_Y$. In the equation, $Y_{\mu\nu}$ is the hypermagnetic field
strength and $\tilde Y_{\mu\nu}=\epsilon_{\mu\nu\alpha\beta}Y^{\alpha\beta}$ is its dual
($\epsilon_{\mu\nu\alpha\beta}$ is the permutation symbol, defined in appendix
\ref{App:Symb}.)

Normally, this source of baryon number violation is ignored because vacuum is
supposed in the initial as well as in the final state.\footnote{
	Even when vacuum is supposed in the initial as well as the final state
	the weak part 
	will still give rise to a baryon
	number changing current \cite{EWBaGe}.
}
However, in the theory of the hypercharge axion, we suppose that a hypermagnetic
field exists before the electroweak phase transition.
Where it comes from would be the issue of another study, but according
to \cite{PriMF} there "are neither compelling theoretical arguments nor motivated
phenomenological constraints which could exclude the existence of magnetic fields
prior to the nucleosynthesis epoch."
The same article also gives the ratio for the baryon-to-entropy ratio $\delta(n_B/s)$
due to the existence of the hyperelectromagnetic fields.
The formula depends on space, time and hypermagnetic field strength.

In \cite{BruOakAmp} it is shown that the existence of a hypercharge axion
introduces a new term in the Maxwell
equations, which is a coupling between the hypercharge axion field $X$ and the
hypermagnetic field $B_Y$.
The solution of the Maxwell's equations can be decomposed in Fourier modes
with different frequencies.
The equation of motion for the hypercharge axion can be solved and
the solution are periodic oscillations or rolling around the minimum
of its potential. 

Depending on the
frequency of the axion oscillation/rolling, it will couple to (=come to resonance with)
different Fourier modes. More specifically there will be a certain mode that will
be maximally amplified, while the nearby modes are amplified somewhat less.
The amplification is exponential
and can achieve values of $10^{12}$ or larger only after a few cycles.
Such high hypermagnetic fields can lead to enough baryon asymmetry generation
through equation~(\ref{eq:divJB}).
In order to preserve the baryon asymmetry, the oscillation or rolling should
take place just before or during the phase transition to avoid diffusion
of the magnetic modes, which erases the field, and thus the conditions for
baryon number violation.
In the scenario of a rolling hypercharge axion, the amplification will
affect very long wavelength modes. These modes do not diffuse as quickly
as higher wavelength modes. This means that the resonance does not have
to occur at the moment of the electroweak phase transition, but could
occur somewhat earlier.


\subsection{The Lagrangian}
The Lagrangian density describing hyperelectromagnetic fields coupled
to the heavy pseudoscalar axion $X$ in the resistive approximation \cite{BruOakAmp}
of the highly conducting electroweak plasma is:
\begin{equation}
\begin{array}{ll}
        L = &\sqrt{-g}\left(\frac 12\nabla_\mu X \nabla^\mu X - V(X) -
	\frac 14 Y^{\mu\nu}Y_{\mu\nu} -J_\mu Y^\mu -\frac \lambda 4 X  Y^{\mu\nu}\tilde Y_{\mu\nu}
	\right)\\\\
	&-\mu \epsilon_{ijk}Y^{ij}Y^k,
\end{array}
\end{equation}
where $Y_{\mu\nu}$ is the hypercharge field strength, 
$Y^k$ is the hypercharge photon,
$g=\det(g_{\mu\nu})$ is the determinant of the metric tensor, $\lambda\sim M^{-1}$
is a scaling constant, $\nabla_\mu$ is
the covariant derivative, $V(X)$ is the axion potential and $J_\mu$ is the Ohmic current. The last term $\mu\ldots$
represents the possibility that a fermionic chemical potential survives the
unbroken phase of the electroweak plasma. Finally, 
$\epsilon^{ijk}$ is the permutation symbol, which is 1 for
all even permutations of $\epsilon^{123}$, -1 for all odd permutations and 0 for
all other values of the indices.

An explanation of the different terms in the Lagrangian can now be appropriate.
\begin{description}
	\item[$\frac 12\nabla_\mu X \nabla^\mu X$]
		is the good old dynamic term, correponding to $\frac{mv^2}2$ in
		classical mechanics.
	\item[$V(X)$]
		is the potential of the hypercharge axion, which can be approximated
		\cite{BruOakAmp}
		to be $\frac 12M_Y^2X^2$, a harmonic oscillator potential.
		$M_Y$ is a mass-scale, see below.
	\item[$\frac 14 Y^{\mu\nu}Y_{\mu\nu}$]
		is the hyperelectromagnetic energy, which is proportional to
		$\vec E_Y~\cdot~\vec H_Y$
		($\vec E$ and $\vec H$ being, respectively, the hyperelectric and hypermagnetic field).
	\item[$J_\mu Y^\mu$]
		is the ordinary current of electric charge.
	\item[$-\frac \lambda 4 X  Y^{\mu\nu}\tilde Y_{\mu\nu}$]
		is the interaction between the hypercharge axion and the hyperelectromagnetic
		field, see below.
\end{description}


\subsection{Couplings}
The part of the Lagrangian that is of interest for the detection in accelerators
is the interaction between the hypercharge axion\footnote{
	In fact, it is this coupling that has given the name to the hypercharge
	axion. The coupling is the same as for the quantum chromodynamics axion,
	which was proposed as a possible explanation for the non-existence of a strong equivalent 
	to the weak $CP$ violating term. For more information, see, e.g.,~\cite{QCDaxion}.
} $X$ and the hyperelectromagnetic field $Y_{\mu\nu}$:
\begin{equation}
\label{eq:LagrAx}
        L = \frac{1}{8M_Y}XY_{\mu\nu}\tilde{Y}^{\mu\nu},
\end{equation}
where $\tilde{Y}^{\mu\nu}=\epsilon^{\alpha\beta\mu\nu}Y_{\alpha\beta}$
is the dual of the hyperelectromagnetic field,
$M_Y=\frac 1 {2\lambda}$ is the mass-scale of the hypercharge axion, which
in principle could be anything that is much larger than the mass of the axion, $m_X$.
We can obtain the Lagrangian in terms of $Z^0$s and photons
if we decompose the hypercharge fields into the $Z^0$, and the photon, 
\begin{equation}
\begin{array}{l}
	Y_{\mu\nu} = \partial_\mu Y_\nu - \partial_\nu Y_\mu,\\
	Y_\mu = A_\mu \cos \theta_W - Z_\mu \sin \theta_W,
\end{array}
\end{equation}
where $A_\mu$ is the photon field, $Z_\mu$ is the $Z^0$, and $\theta_W$ is the weak mixing angle.
We obtain 
\begin{equation}
\label{eq:Coupl}
\begin{array}{ll}
	L = \frac {X\epsilon^{\mu\nu\rho\sigma}}{8M_Y}[
		&a (\partial_\mu A_\nu\partial_\rho A_\sigma)
		+b (\partial_\mu Z_\nu\partial_\rho Z_\sigma)\\
		&-c (\partial_\mu A_\nu\partial_\rho Z_\sigma)
		-c (\partial_\mu Z_\nu\partial_\rho A_\sigma)
		],
\end{array}
\end{equation}
where $a =  \cos^2\theta_W$, $b=\sin^2\theta_W$ and $c=\sin\theta_W\cos\theta_W$.
Here we can clearly see the interactions and their coupling constants.
In the first term we see the $X\rightarrow \gamma\gamma$ coupling,
in the second the $X\rightarrow Z^0Z^0$ coupling, and in the third and fourth
the $X\rightarrow Z^0\gamma $ coupling.


\subsection{Branching Ratios}
\label{sec:BR}
The \gloss{branching ratios} of the processes can be obtained from the matrix elements
of the Lagrangian:
\begin{equation}
\begin{array}{l}
	\Gamma_{X\rightarrow \gamma\gamma} = \frac 1{32\pi M_Y^2}\cos^4\theta_W\left[m_X^3\right]\\
	\Gamma_{X\rightarrow Z^0\gamma} = \frac 4{32\pi M_Y^2}\cos^2\theta_W\sin^2\theta_W\left[
		\frac{(m_X^2-m_Z^2)^3}{M_X^3}\right]\\
	\Gamma_{X\rightarrow Z^0Z^0} =  \frac 1{32\pi M_Y}\sin^4\theta_W\left[(m_X^2-4M_Z^2)^{3/2}\right],
\end{array}
\end{equation}
where $\theta_W$ is the weak mixing angle, $m_X$ is the mass of the axion,
$M_Y$ is the mass-scale, and $M_Z$ is the $Z^0$ mass.

We note that the squared coupling constants from equation~(\ref{eq:Coupl}) are here
as well as the mass-scale squared. The factor 4 for $X\rightarrow Z^0\gamma$ comes
from the two last terms in equation~(\ref{eq:Coupl}) which are identical.

The total width for $m_X\sim 1$ TeV, $M_Y\sim 1$ TeV is about 1 GeV.

\subsection{Candidates for the Hypercharge Axion in Extensions of the Standard Model}
\label{sec:Extensions}
Brustein and Oaknin treat this subject in \cite{BruOakExt}. 
There are several pseudoscalars in different extensions of the standard model
that couple to the hypercharge as proposed in equation~(\ref{eq:LagrAx}).
In order to drive baryogenesis, a condition on the rolling/oscillating time
must be fulfilled \cite{BruOakExt}.

One candidate would be the heavy Higgs pseudoscalar in the supersymmetric standard model,
but this coupling is shown to vanish in the symmetric phase of the electroweak theory.
Other possible candidates, like the pseudo\-scalar component of sneutrinos, 
do not couple to the hypercharge.

In string theory there are several possible pseudoscalars but the condition mentioned
above turns out to be difficult to fulfil. Besides, the coupling to the hypercharge
generally happens at temperatures much higher than the electroweak phase transition.
In summary, it is possible that the hypercharge axion could be described by 
string theory, but the conditions are rather severe.


\chapter{Phenomenology of the Hypercharge Axion}

\label{ch:CaEx}
This chapter discusses the experimental aspects and the phenomenology
of the hypercharge axion. It describes the signatures and background
for the detection of such a particle in the ATLAS detector at the future large hadron collider (LHC) at CERN.


First the detector is presented briefly to situate the project in the context.
Second, the characteristics of axion decay are discussed as well as the process that
is the subject of this thesis
\begin{equation}
        q + \bar q \rightarrow q + \bar q + \gamma + \gamma \rightarrow q + \bar q + X
\end{equation}
with subsequent decay of the axion to photons and $Z^0$s.
Third, the approximations are presented. The interactions are simulated with
a Monte Carlo program (PYTHIA) and the detector with ATLFAST. The semi-classical
Weizs\"acker-Williams approximation is made for the emission of photons from
the primary partons.
Fourth, the general backgrounds ($\gamma,Z^0$ and jets) are discussed with their
characteristics and their effect on our signals.
Fifth, the different processes that originate from the existence of
the axion are presented. There are several, but only a few can be detected in ATLAS.

\section{The Detector}
\subsection{Large Hadron Collider}
The large hadron collider (LHC) is a particle accelerator that is built at CERN
(European laboratory for particle physics) and will be ready for experiment in 2005. The accelerator will bring
protons and, later, ions into head-on collisions with higher energies than ever before.
These high energies will allow physicists to probe deeper into the structure and the
fundamental properties of matter and energy. The elevated energy will allow new
types of particles to be created, detected and measured, but they will not be
easy to detect.

Along with the new, interesting physics there will be huge amount
of background, or noise, coming from other processes that are already known.
The principal reason behind this background is the fact that a proton is made
of three quarks, which makes the collision a six-body problem. It is not
easy to calculate and predict theoretically, nor to accurately detect all
outgoing particles.

LHC will be able to produce three basic types of reactions.
	First, it can collide
proton beams, each proton having an energy of 7 TeV, for a total of 14 TeV
in the interaction. This will be the main type of collision at LHC and
the luminosity\footnote{
	Luminosity $\cal L$, is directly proportional to the number of particles
	in each bunch and to the bunch-crossing frequency, and inversely
	proportional to the area of the bunches at the collision point.
} will reach $10^{34}$ cm$^{-2}$s$^{-1}$, which is about
100 times higher than that of current colliders.
	Second, heavy ions, like lead, can be collided with a total
energy of 1250 TeV in the collision, about 30 times higher than that of the
Relativistic Heavy Ion Collider (RHIC) at the
Brookhaven Laboratory in the US.
	Third, on long term it is planned to combine LEP and LHC. Electron-proton collisions can be obtained
with total energies of $\sim 1.5$ TeV, about five times higher than HERA
at the DESY laboratory in Hamburg.

\subsection{The ATLAS detector}
The ATLAS (A Toroidal LHC ApparatuS) detector is one of the detectors on LHC.
It is immense (like a five-story building) and is hermetically closed.
The parts of ATLAS that are relevant for this research are:
\begin{description}
	\item[Inner tracker] -
		measures the momentum of each charged particle.
		The inner tracker is inside a solenoidal magnet of 2 Tesla,
		bending the charged particle tracks and allowing measurement of their
		momenta.
	\item[Calorimeter] -
		measures the energies carried by the particles.
		There is one electromagnetic calorimeter and one hadronic calorimeter.
		The calorimeters are thick enough to let through only $10^{-7}$ percent
		of the entering particles.
	\item[Muon spectrometer] -
		identifies and measures muons.
		The muons interact so feebly with matter that most of them pass
		the calorimeter without stopping and without causing a shower
		of particles. The muon spectrometer 
		consists of chambers to measure their track.
		A strong toroidal magnetic field is present to measure their momenta.
\end{description}
There is also a trigger system that selects hundreds of interesting events from
thousands of millions of others and a data aquisition and filtering system to further
classify and distinguish each event (collision every 25 ns). 
	
Obviously, the detector has a limited resolution, which, e.g.,~means that
a quark will be observed as a shower, a jet. Also, photons
with low transverse momentum will not be detected, and 
particles have to be separated by a certain angle to be identified as
different particles.

\section{Characteristics of the Hypercharge Axion}
\subsection{Interactions}
What is interesting for us is the interaction the axion would have with
matter. Let us investigate 
the interaction of the pseudoscalar axion field $X$ with the
hypercharge field strength $Y_{\mu\nu}$. This is described
by the Lagrangian
\begin{equation}
	L = \frac{1}{8M_Y}XY_{\mu\nu}\tilde{Y}^{\mu\nu},
\end{equation}
where $M_Y$ is a scale factor with the dimension of energy,
and $\tilde{Y}^{\mu\nu} = \epsilon^{\alpha\beta\mu\nu}Y_{\alpha\beta}$
is the dual of the hypercharge field strength. This is the Lagrangian
proposed by Brustein and Oaknin
\cite{BruOakColl}. Obviously there are several other interactions in the
system, including potentials and self-interactions, but these do not
influence the results of this section.
For a more detailed discussion, see~section \ref{sec:Axion}. 
We will take $M_Y$ as a free parameter and determine for
which values of $M_Y$, and for which masses of the axion $m_X$,
it will be possible to discover the axion in ATLAS . The actual
values of $M_Y$ and $m_X$
depend on the theory invoked to make the connection between
the matter-antimatter problem and the hypercharge axion, see~section \ref{sec:Extensions}.

The hypercharge field-strength consists basically of $Z^0$s and photons,
which means that the interaction will be between the axion and either two photons,
two $Z^0$s or one $Z^0$ and one photon. Hence, to identify the hypercharge
axion in the detector, we will look for these three types of decay products.
With the particles coming out from the collision, we will look for photons
and\footnote{
	In fact, we look for the decay products of the  $Z^0$s, see below.
} $Z^0$s and reconstruct the invariant mass\footnote{
The invariant mass $M$ is defined as
\[
	\textstyle{M^2 = (\sum_iE_i)^2 - \left|\left |\sum_i\vec p_i\right|\right|^2,}
\]
where $E_i$ and $\vec p_i$ are, respectively, the energy and the momentum
of particle $i$, and $||\vec v||^2 = v_x^2 + v_y^2 + v_z^2$ is the squared norm
of the vector $\vec v$.
} of these particles. This mass should be equal to the mass of the axion,
and with a lot of events it should be recognizable above the background.

As the hypercharge axion is expected to be rather heavy, its decay products will often
have a high transverse momentum. Most other particles will more or less
follow the direction of the beam due to the high momentum in the beam
direction. Hence, another characteristic of the axion will be off-axis
photons and $Z^0$s.

\subsection{The Event}
Brustein and Oaknin \cite{BruOakColl} proposed the channel
$f+\bar f \rightarrow Z^*/\gamma^* \rightarrow Z/\gamma X$
for detection of the hypercharge axion $X$. However, the cross section
for this process is rather low and another possibility
to detect the axion will be investigated here to complement their work.

\begin{figure}
        \centering
        \includegraphics[width=3in]{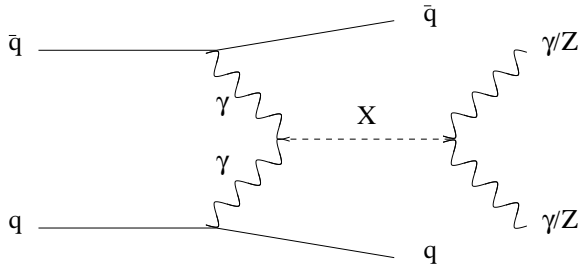}
        \caption{The event that is the object of this study. Two
		quarks within the proton emit two photons, which
		collide and produce the axion, which disintegrates
		to photons and/or $Z^0$s.}
        \label{fig:Event}
\end{figure}

We will study emission of two photons (one from each proton) colliding to give the axion.
In the proton, generally only one of the quarks (an up or a down quark) or a gluon will interact
in the collision with another quark or gluon from the other proton.
The event that is the subject of this research can therefore be written as
(figure \ref{fig:Event})
\begin{equation}
	q + \bar q \rightarrow q + \bar q + \gamma + \gamma \rightarrow q + \bar q + X,
\end{equation}
where the axion can disintegrate as
\begin{equation}
\begin{array}{ll}
	X &\rightarrow \gamma + \gamma\\
	& \rightarrow Z^0 + \gamma\\
	& \rightarrow Z^0 + Z^0.
\end{array}
\end{equation}

We can make a rough estimate of the number of events 
for a certain mass of the axion, say 1 TeV.
The number of events is, by definition, the product of luminosity, time
and cross section:
\begin{equation}
\label{eq:Nev}
	N_{ev} = {\cal L}\times t\times \sigma \sim
	10^{28} \textrm{ cm}^{-2}\textrm{s}^{-1} \times
		10^7 \textrm{ s} \times
		150 \textrm{ pb} \times
		10^{-24} \textrm{ b}/\textrm{cm}^2
	\approx 15.
\end{equation}
Here the $\gamma\gamma\rightarrow X$ cross-section, $\sigma$,
at the resonance is taken from \cite{BruOakColl} with the mass-scale
$M_Y=1$ TeV, and where the luminosity
$\cal{L}$ is extrapolated from the graph in \cite{PapaHiggs}
and integrated over the resonance width of the axion (about 1 GeV). This is the luminosity
for the production of two photons with invariant mass $\sim 1$ TeV during
LHC operation at low luminosity. 
It is calculated assuming coherent emission from a proton.
In fact, it can be much higher if the photons are emitted from
quarks and the proton is allowed to disintegrate. The time $t$ is taken to be one year of running, or $10^7$ s,
and $10^{-24} \textrm{ b}/\textrm{cm}^2$ is a rough conversion factor from cm$^{-2}$ to barn.
This is detectable, though the signal will be rather feeble. This means
that we have to be more careful with the background. To increase the signal we
will assume that we run during three years of high luminosity at LHC, giving
ten times as many events:
\begin{equation}
	\label{eq:ApprNev}
	N_{ev} \sim 150.
\end{equation}

The forward jets could in principle serve as a signature for the process, but
the enormous jet background in the LHC will completely destroy this possibility.
This jet background (also called QCD background) comes from interactions between quarks.
These interactions produce outgoing quarks that
will give showers of particles in the decay process. These showers are detected
in ATLAS and are called jets. Because LHC works with proton collision (i.e., strong interactions) at
extremely high energies there will also be a large number of jets,
in the background.

We can also use the fact that the decay products of
the axion (the $Z^0$s and photons) are produced virtually \gloss{back-to-back} in the transverse
plane, i.e., the angle between the two particles in the transverse plane is 180 degrees.
The reason for this is that the two photons producing the axion have almost
zero transverse momentum (in comparison with their longitudinal momentum)
which comes from the fact that the quarks in the proton
have very low transverse momenta.

In summary, the characteristic of the axion are
	decay products $X\rightarrow \gamma\gamma$, $Z^0Z^0$ or $Z^0\gamma$
	with an invariant mass producing a bump, emitted
	back-to-back and with 
	high $p_T$.


\section{Approximations}
\subsection{Monte Carlo Simulation}
The events have been generated in PYTHIA \cite{PYTHIA} which is a Monte-Carlo simulation
program for particle interactions. To simulate the detector effects with its limits
in resolution and acceptance ATLFAST has been used. This means, among other things, that
photons with $p_T<60$ GeV will not be detected and that photons which
are too close to the beam axis will be mistaken for jets.



The generation of the matrix elements to calculate the interactions
have been provided by Gilles Couture. He has used
 the Weizs\"acker-Williams approximation for the emission of the two
photons and 
the parton distribution functions from \cite{PDF}, mostly Appendix A.

\subsection{Weizs\"acker-Williams Approximation}
For the interaction between the two quarks, the \gloss{Weizs\"acker-Williams approximation}
(also called equivalent photon approximation or almost-real-photon approximation)
has been made. 
In principle the electromagnetic
field of the charged particle is treated as (almost) real photons. These can
interact with the almost real photons from the other charged particle.
This approximation is only valid when the transverse momentum of the charged particle is
virtually zero, and when they have ultra-relativistic speeds.
This should be fairly well accomplished in our scheme, as the charged
particle is a quark, and the quarks inside a proton have very little transverse
momentum and the speed is only a small fraction from that of light.
Another consequence of this approximation is that the protons are supposed to split
and produce fragments in the forward direction.

There does not exist any good theoretical description of $\gamma\gamma$ processes from proton-proton collisions.
However, the Weizs\"acker-Williams approximation suffices well.


%



\section{Background}
The background consists of all processes that give the same type of signal as
the interesting event. An everyday example would be to weigh different animals, trying
to distinguish them {\it only} by their weight.
Suppose that we look for the elephant (the axion). We would have an enormous
number of light animals, like insects, indistinguishable from each other, (loosely
corresponding to the QCD background, see below). As the size goes up,
fewer and fewer animals are heavy enough. Eventually, we might see a bump for
humans around 70 kg, and around 5 tons we find the indian elephant bump. However, there
are other animals that could be this heavy, rhinoceros and whales for example, though it would
be rare. This is the background. 

This section is dedicated to the different backgrounds for the axion. First, the
interesting processes will be presented so that we know what backgrounds to consider.
In summary, these backgrounds are processes producing photons and $Z^0$s each of which will
be discussed separately. A brief discussion of the jets is also included.

In the calculation of the backgrounds for the different processes, the contribution 
of the Higgs boson is excluded. This is done because the Higgs boson has
an unknown mass, which makes it difficult to 
estimate the background it could produce for the process.
In fact, we do not even know that it exists.
If it exists as expected with a mass of 115 GeV, its contribution
would be negligible, because the branching ratios of $H^0$ to photons
and $Z^0$s are very small \cite{Higgs}.

\subsection{Decay Channels}
The possible decay channels of the axion are (with branching ratios\footnote{
	The branching ratios of the axion depend on its mass,
	but this dependence is very weak (see~section \ref{sec:BR}).
	In the range $500<m_X<5000$ GeV
	the difference is less than a few tenth of percentage.
} over the arrows):
\begin{equation}
\begin{array}{lllll}
	X &\xrightarrow{44\%}&\gamma +\gamma&\\
	&\xrightarrow{52\%}&\gamma + Z^0& \xrightarrow{6\%}& \gamma + l + \bar l\\
	&&& \xrightarrow{70\%}& \gamma + jet + jet\\
	&\xrightarrow{4\%}& Z^0 + Z^0& \xrightarrow{0.36\%}& l + \bar l + l + \bar l\\
	&&& \xrightarrow{3.6\%}& l + \bar l + jet + jet\\
	&&& \xrightarrow{36\%}& jet + jet + jet + jet,
\end{array}
\end{equation}
where we have used the fact \cite{PDG} that the $Z^0$ decays to
two specific leptons 3 percent of the time, to two quarks (jets) 70 percent of
the time and to neutrinos the rest of the time. The $\tau$ is more
difficult to detect at the LHC and is not considered here,
leaving the electron and muon, for a total
of 6 percent detectable lepton decays.



As we can see, our primal branchings will be:
\begin{equation}
\begin{array}{lll}
	X &\rightarrow \gamma +	\gamma&\\
	X &\rightarrow \gamma +	Z^0 \rightarrow \gamma + l + \bar l
\end{array}
\end{equation}
because $X \rightarrow Z^0 + Z^0 $ is rare and $X \rightarrow \gamma + jet +jet$
is very polluted by the QCD background. A more detailed discussion follows below in
section \ref{sec:Proc}. Before discussing this further some general backgrounds
will be described.

\subsection{$\gamma$ Background}
We can have photon emission at different stages in the calculations.
First, in \gloss{initial and final state radiation}. Second, in specific
processes giving photons as end-products. Third, a jet can be taken
to be a photon in the detector.

\subsubsection{Initial and Final State Radiation}
Initial and final state radiation are decays of particles that
occur before and/or after the event.
Examples of initial/final state radiation are $q \rightarrow q\gamma$ and
$q \rightarrow qg$.

\subsubsection{Photon Processes}
\label{sec:Photons}
The principal processes that give rise to photon production in LHC are
\begin{equation}
\label{eq:PhoBg}
\begin{array}{llllll}
        f\bar f      &\rightarrow&   \gamma/Z^0		&       fg      &\rightarrow&   f + \gamma\\
        f\bar f      &\rightarrow&   g + \gamma		&	gg      &\rightarrow&   g + \gamma\\
        f\bar f      &\rightarrow&   \gamma + \gamma	&       gg      &\rightarrow&   \gamma + \gamma\\
        f\bar f      &\rightarrow&   \gamma + Z^0,	&&&
\end{array}
\end{equation}
where the fermion $f$ is a quark in our case and $g$ is the gluon.
The processes
$f\bar f      \rightarrow   g + \gamma$ and
$gg      \rightarrow   g + \gamma$ have a high cross-section,
but on the other hand they do not correspond to the exact final
state of any of our processes. As for the other processes,
they all have a very small cross-section, leaving us with
a fairly small photon background.

Another process that could be expected to contribute is $gg\rightarrow \gamma Z^0$.
The contribution would be to the background of $X\rightarrow Z^0\gamma$,
but it is not included in PYTHIA and will not be included in our study.


\subsubsection{Fake Photons}
\label{sec:FalsePho}
The resolution of the photons is rather good, but in some rare cases (approximately 1 time in 3000)
a jet can be mistaken for a photon \cite{TDR}. When the signal is strong, this is not a problem.
However, when the signal is weak,
 this
"fake" photon background could also be significant.

\subsection{$Z^0$ Background}
\label{sec:BgZ0}
As two of the three branchings of the axion produce $Z^0$s it is important to see
which other processes give $Z^0$s.
A $Z^0$ can be produced in several different ways:
\begin{equation}
\label{eq:ZBg}
\begin{array}{llllll}
        f+\bar{f} &\rightarrow & \gamma/Z^0	&	f+g &\rightarrow & f + Z^0\\
        f+\bar{f} &\rightarrow & g + Z^0	&	(Z^0Z^0  &\rightarrow&   Z^0 + Z^0)\\
        f+\bar{f} &\rightarrow & \gamma + Z^0	&	(W^+W^-      &\rightarrow&   Z^0 + Z^0)\\
        f+\bar{f} &\rightarrow & Z^0 + Z^0
\end{array}
\end{equation}

Among these processes $f+\bar{f} \rightarrow  g + Z^0$ and $f+g \rightarrow  f + Z^0$
have high cross-section but do not directly lead to one of the final states
of the axion decay either. This means that the missing photon will have to
come from initital and/or final state radiation, as described in section \ref{sec:Photons}.
Fortunately, a high-$p_T$ photon from these sources is very improbable, diminishing
the importance of these background processes.

For $Z^0Z^0  \rightarrow   Z^0 + Z^0$ and $W^+W^-      \rightarrow   Z^0 + Z^0$,
the cross-sections are extremely small, making it rather safe to exclude them
from the calculations (see~equation \ref{eq:Processes}).


\subsection{Jet Background}
\label{sec:JetBg}
The QCD processes that have been generated are
\begin{equation}
\begin{array}{llllll}
	q_iq_j		&\rightarrow&	q_iq_j	&	q_i\bar q_i	&\rightarrow&	q_k\bar q_k\\
	q_i\bar q_i	&\rightarrow&	gg	&	q_ig		&\rightarrow&	q_ig\\
	gg		&\rightarrow&	q_k\bar q_k&	gg		&\rightarrow&	gg,
\end{array}
\end{equation}
where the evolution and showers come from initial and final state radiation
as described in section \ref{sec:Photons}.

In LHC there will always be an enormous background from jets due to
the high energy and luminosity of the beam. This means that it will
be very noisy to try to detect events with jets as final states.
If the mass of the axion is relatively high, the criterion of
high-$p_T$ decay products can be used to reduce the background
significantly. Normally, the produced jets will follow the beam
direction because the quarks have relatively low-$p_T$ in the protons.
However, the problem with a
heavy axion is that its cross section for axion production decreases with increasing mass.
This means that even with a low background, the cross-section for
axion production might be too low to allow detection.

We know that the signal studied should have two and only two jets. However, since
we have used the Weizs\"acker-Williams approximation we cannot exploit this characteristic
to apply cuts on the number and angle of the jets.
\section{Processes}
\label{sec:Proc}
As can be seen from
the branching ratios in the previous section and the estimate of number
of events, equation~(\ref{eq:ApprNev}), there will be almost no events with production
of the $Z^0$s. Furthermore, LHC has a huge QCD background (jets) making
it a difficult task to discern $Z^0 \rightarrow jet + jet$.
This leaves us with two principal processes, $ X \rightarrow \gamma + \gamma $
and $X \rightarrow \gamma + Z^0 \rightarrow \gamma + l + l$.
These will be described in more detail below, while the others will be treated
briefly in a separate subsection.

\subsection{Process $ X \rightarrow \gamma + \gamma $}
This process should be very clean. Photons are rather easy to detect and
their energies and momenta can be determined with good precision. 
Most of the noninteresting events (background) will be possible to
sort out because the photons are produced back-to-back in the transverse plane,
generally with high transverse momentum.

The background processes that should be important are
\begin{equation}
\begin{array}{l}
	f\bar f \rightarrow \gamma\gamma\\
	gg \rightarrow \gamma\gamma,
\end{array}
\end{equation}
along with the other processes in equation~(\ref{eq:PhoBg}) with initial and final
state radiation to produce the other photon.

The $\gamma\gamma$ decay channel will be our principal source of events
that hints at the existence of the axion
because of its high branching ratio (44 percent) and the fact that photons are easily
identified and (normally) do not decay.

\subsection{Process $X \rightarrow \gamma + Z^0 \rightarrow \gamma + \bar l + l$}
This process will also be very clean, but unfortunately also very rare.
The total branching ratio for this process will be 3.1 percent and with the estimated
total number of events equation~(\ref{eq:ApprNev}) this leaves about five events
that will be produced during three years of high luminosity at the LHC.
When we add the limits of the detector, along
with the condition that three particles should be detected ($\gamma \bar l l$),
the number
of observed events should be very small, at least for $m_X\sim 1000$ GeV.

The background processes that should be important are
\begin{equation}
\begin{array}{l}
	f\bar f \rightarrow \gamma/Z^0\\
	f\bar f \rightarrow \gamma+Z^0,
\end{array}
\end{equation}
and the other processes in equation~(\ref{eq:ZBg}) with initial and final state
radiation providing the photon.

\subsection{Other processes}
\label{sec:OthProc}
\subsubsection{Process $X \rightarrow \gamma + Z^0 \rightarrow \gamma + jet + jet$}
The process $X \rightarrow \gamma + Z^0 \rightarrow \gamma + jet + jet$ has
a fairly high branching ratio (36 percent), which gives a good number of such events.
However,
as mentioned before,
the fact that the jets have to be used to reconstruct the invariant
mass of the $Z^0$ is very unfortunate. Even with the back-to-back criterion
and the high-$p_T$ criterion, the QCD background (see~section \ref{sec:JetBg}) is overwhelming and
effectively drowns the signal.

\subsubsection{Process $X \rightarrow Z^0 + Z^0 \rightarrow l + l + l + l$}
The process  $X \rightarrow Z^0 + Z^0 \rightarrow l + l + l + l$ should be very clean, but also very rare.
Combining the low total branching ratio (0.0144 percent) with the difficulty for the detector to observe four
distinct leptons leaves us without a single such event during three years of high
luminosity at the LHC.

Still, the event will be very clean because 
each pair of leptons can be combined to produce the invariant mass of $Z^0$. The
$Z^0$s then give the invariant mass of the axion and, in addition, they should be back-to-back,
like in the case of $X \rightarrow \gamma \gamma$.

Possible background events come from $f\bar f\rightarrow Z^0Z^0$.

\subsubsection{Process $X \rightarrow Z^0 + Z^0 \rightarrow l + \bar l + jet + jet$}
This process should be rather polluted because of the large QCD background that
will give jets, and $Z^0$ emission that will give the leptons, as discussed in section
\ref{sec:BgZ0}.
Furthermore, the signal in itself will be very weak (branching ratio 0.14 percent) and difficult
to detect with its four particle final state.

An example of background is
\begin{equation}
	t + \bar{t} \rightarrow b W^+ + \bar b W^- + \rightarrow jet+jet + l^+ + \nu + l^- +\bar\nu.
\end{equation}

\subsubsection{Process $X \rightarrow Z^0 + Z^0 \rightarrow jet + jet + jet + jet$}
The process $X \rightarrow Z^0 + Z^0 \rightarrow jet + jet + jet + jet$
has only the back-to-back and high-$p_T$ criteria to differentiate it from
the enormous QCD background. These criteria are far too weak to make the signal discernable
above the background.

\chapter{Results and Analysis}
\label{ch:ReAn}
As discussed in the previous chapter, the simulation of the processes and
the backgrounds has been done in PYTHIA \cite{PYTHIA}, with the axion interaction as
an externally included process and with ATLFAST for simulation of the ATLAS
detector.

This chapter consists of four parts: The simulation of the signal of
the different decay channels, the simulation of the background, 
both together, and finally some conclusions. The simulation of
the signal is done for different masses of the axion and gives us
a hint of where a search could be fruitful. The signal will be stronger
for lower masses, but the background will also be stronger. The analysis
of the background will be done completely for $m_X\sim$ 1000 GeV. This gives 
an indication of which background processes are significant. This is
important to know because at lower energies we have to generate more
events as the cross-section goes up. The backgrounds that are found
to be important are simulated also for $m_X\sim 800$ GeV.


\section{Signal}
\begin{figure}
	\begin{minipage}[t]{0.45\textwidth}
		\begin{center}
		\mbox{\epsfig{file=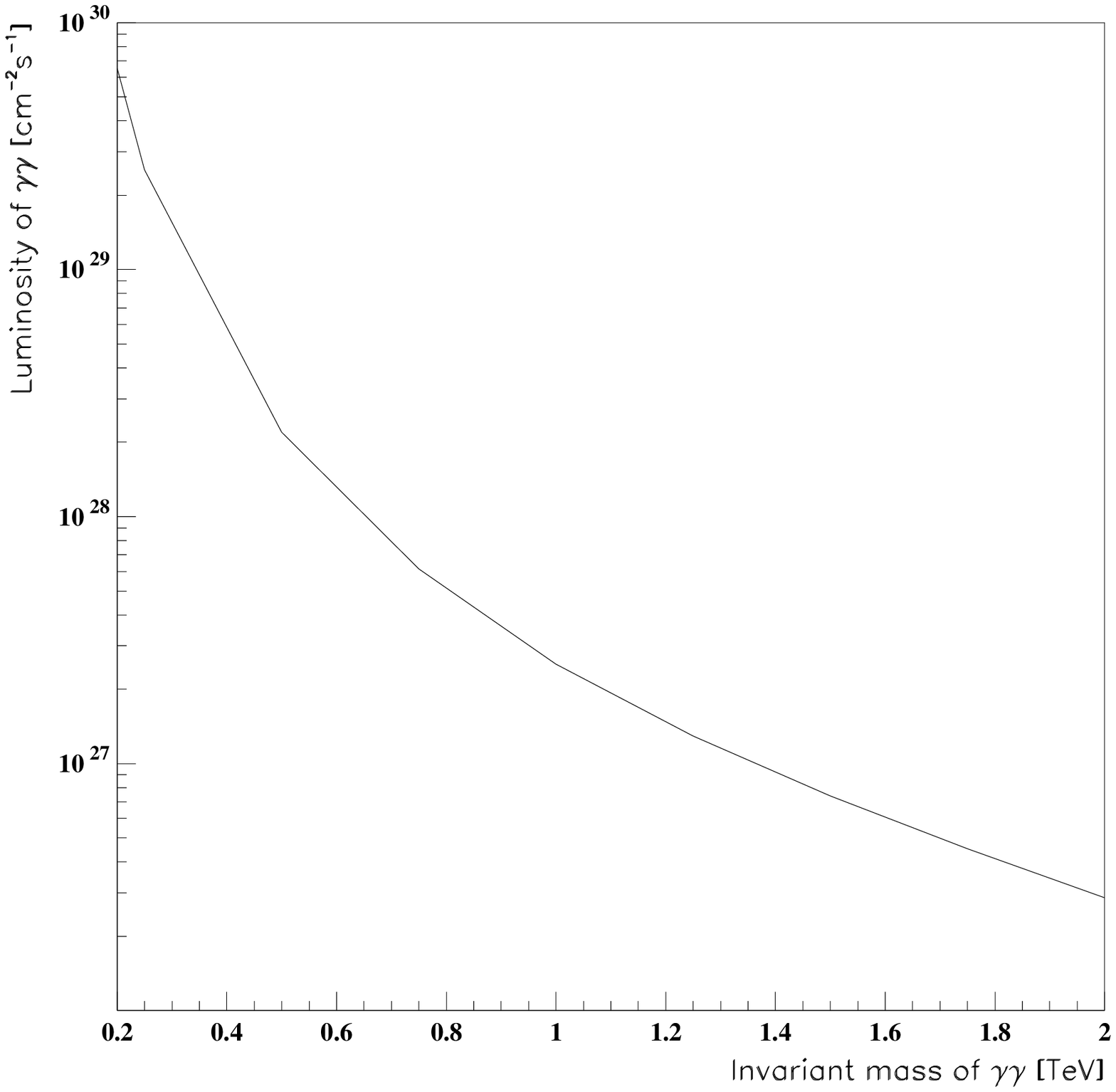,width=6cm}}
		\hspace{6cm}
		\end{center}
		\caption{Luminosity of $\gamma\gamma$ production
			as a function of their invariant mass for
			a pp luminosity of $10^{33}$ cm$^{-2}$s$^{-1}$.}
		\label{fig:Lum}
	\end{minipage}
	\hfill
	\begin{minipage}[t]{0.45\textwidth}
		\begin{center}
		\mbox{\epsfig{file=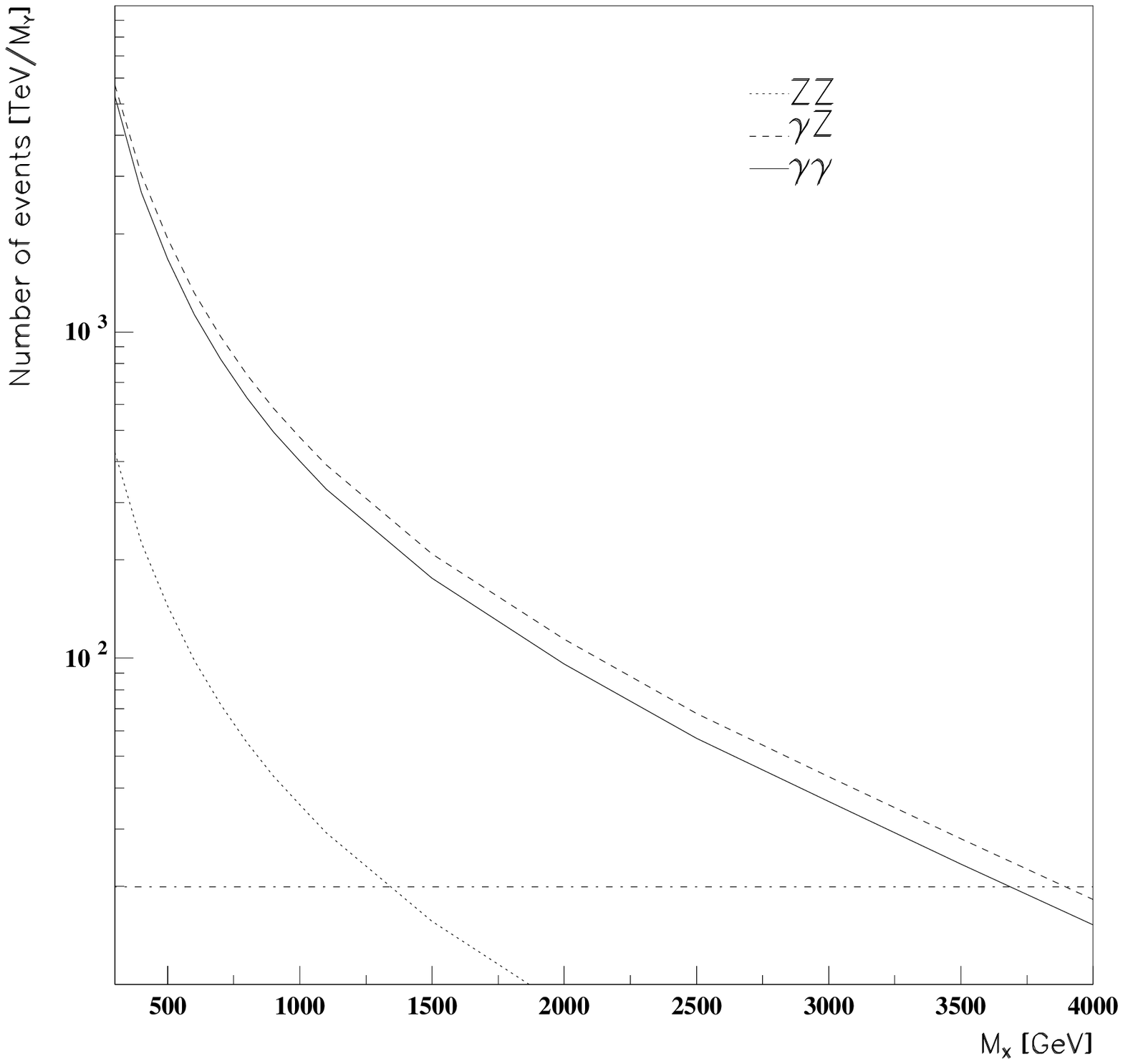,width=6cm}}
		\end{center}
	\caption{Number of events produced for the different processes for three years
		of high-luminosity LHC operation.}
	\label{fig:Nev}	
	\end{minipage}
\end{figure}

First of all we calculate the luminosity for different invariant masses of $\gamma\gamma$.
The process is $q+\bar q\rightarrow q+\bar q\gamma\gamma$. These are the photons that
will interact to produce the axion and consequently their invariant mass should
be the invariant mass of the axion.
When we assume a total luminosity of $10^{33}$ cm$^{2}$s$^{-1}$ we obtain the
result shown in figure \ref{fig:Lum}.
Now, we calculate the cross-sections of the processes.
With the cross-sections we calculate how many events we expect through
\begin{equation}
	N_{ev} = {\cal L}\times t\times \sigma.
\end{equation}
We will require at least ten events and we assume three years of high 
luminosity, which gives an integrated luminosity\footnote{
	Integrated luminosity is ${\cal L}\times t$.
} of 300 fb$^{-1}$.
The number of events will be measured in units of $(\frac{\textrm{TeV}}{M_Y})^2$.
However, we lose more than half of the events in the ATLAS detector due to
problems with resolution and the fact that the
detection of photons is not possible at near forward angles.
This means that we need at least $N_{ev}> 20$ to really detect
the particle. Figure \ref{fig:Nev} shows the number of axions produced as a function of axion mass,
$m_X$, for the different decay channels.
We see that for an axion mass $m_X\sim 1000$ GeV the number of events is almost
a factor ten higher than the one predicted from equation~(\ref{eq:ApprNev}). The reason
for this is that the photon emission is done directly from the quarks, and 
the proton is allowed to disintegrate (see~explanation following equation~(\ref{eq:Nev})).
Note that this is only a first estimate and that the background has to be considered
too. This is done in the following section.

%
%

\section{Backgrounds}
\label{sec:Bgs}
In this section we will discuss preselection cuts on the background, which means that
we limit the kinematic phase space in order to decrease the cross-section. Otherwise we would
have to generate an enormous amount of background. We also generate the complete backgrounds
for $m_X\sim 1000$ GeV and the significant backgrounds for $m_X\sim 800$ GeV.
The signal and the backgrounds are discussed together in the next section, \ref{sec:BgSig}.

\subsection{Cuts}
\label{sec:cuts}
In order not to a have to generate an overwhelming number of background events we
introduce preselection in the calculations. A preselection means that an event is rejected at
an early state because it does not fulfil certain criteria. The cuts in
our simulation have been made on the transverse momentum $p_T$ and the
center-of-mass energy $\sqrt{\hat s}$ of the two partons participating in the
reaction. As discussed in chapter \ref{ch:CaEx}
a characteristic of the axion is high-$p_T$ decay products, motivating a cut
exclude $p_T< 300$ GeV which should be valid for $m_X > 700$ GeV. In order to
even create something similar to the axion, the center-of-mass energy should
be comparable to the mass of the axion. We have chosen to exclude
$\sqrt{\hat s} < m_X-100$ GeV.

\begin{figure}
        \centering
        \includegraphics[width=3in]{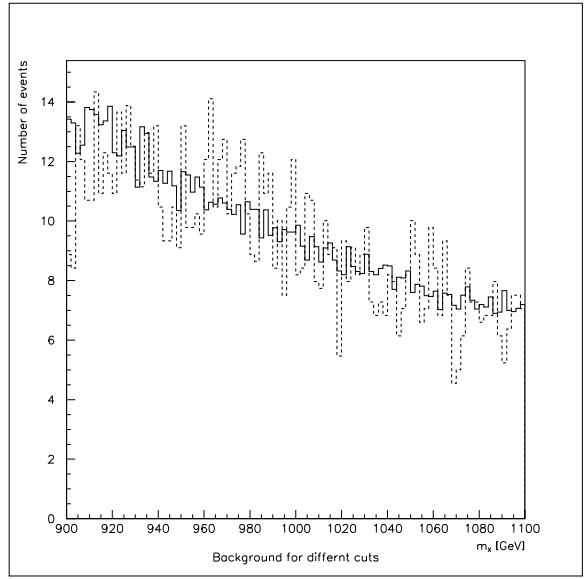}
        \caption{Comparison of reconstructed $m_{\gamma\gamma}$ from background 
		$        f \bar f        \rightarrow   \gamma + \gamma$,
		with preselection cuts, $\sqrt{\hat s}> 900$ (full histogram)
		and $\sqrt{\hat s}> 700$ (dashed histogram).}
        \label{fig:cmpcuts}
\end{figure}

To validate these cuts, the backgrounds have been simulated with lower
cuts. With proper normalization\footnote{
	When two signals have different cross-sections they must be normalized.
	The normalization factor is $N = \frac {\sigma_1/N_1}{\sigma_2/N_2}$,
	where $N_i$ is the number of generated events for signal~$i$.
} the two signals are then compared. In the region around $m_X$ the shape
of the curves should be similar. An example is shown in figure \ref{fig:cmpcuts}.

These were preselection cuts, which act in the generation of events
to reduce the cross-sections. In consequence of the 3$^{\textrm{rd}}$ characteristic
of the axion we apply a third cut, but this can only be done after
the generation of events. This is a cut on the angle in the transverse plane between the particles
that could have been the decay products of the axion. How these particles are
selected is discussed below. The cut on the angle is $3.0<\phi<\pi$.

Obviously, all these cuts must also be applied on the signals themselves.
This will reduce the signal, but even more so the background, which in principle
should be randomly distributed.

\subsubsection{Selection Criteria}
\label{sec:SelCrit}
The selectrion criteria for the different processes are:
\begin{description}
	\item[$X\rightarrow \gamma\gamma$] 
		Select the two photons with highest transverse momentum
		and calculate their invariant mass.
	\item[$X\rightarrow Z^0\gamma\rightarrow \bar ll\gamma $]
		Select the two leptons with highest transverse momentum,
		calculate their invariant mass (should give the $Z^0$),
		select the photon with highest transverse momentum
		and calculate the invariant mass with the reconstructed $Z^0$.
	\item[$X\rightarrow Z^0\gamma\rightarrow jet + jet + \gamma $]
		Same as $Z^0 \rightarrow ll$ except that the leptons
		are replaced by the jets.
	\item[$X\rightarrow Z^0 Z^0\rightarrow \bar ll\bar ll$]
		Select the four leptons with highest transverse momentum,
		calculate the invariant masses between all the leptons,
		choose the pair that gives the mass closest to that of the $Z^0$.
		The other pair is supposed to give the other $Z^0$, and together the pairs
		give the mass of the "axion".
	\item[$X\rightarrow Z^0 Z^0\rightarrow \bar ll+jet+jet$]
		Same principle as for $Z^0 Z^0\rightarrow \bar ll\bar ll$.
	\item[$X\rightarrow Z^0 Z^0\rightarrow jet+jet+jet+jet$]
		Same principle as for $Z^0 Z^0\rightarrow \bar ll\bar ll$.
	\item[QCD background]
		In some rare cases (about 1 in 3000) a jet can be mistaken for
		a photon and the selection process for this is to choose
		the photon and the jet with highest $p_T$ and calculate
		their invariant mass.
\end{description}

\subsection{Complete Background for $m_X\sim 1$ TeV}
\label{sec:mX1000}
As stated in the previous chapter, there are several different
backgrounds that will affect the signal. 
The cross-sections are calculated with the two initial cuts, $p_T>300$ GeV
and $\sqrt{\hat s}>900$ GeV. For three years of high luminosity at the LHC, the integrated luminosity is 300 fb$^{-1}$
and the number of expected events is calculated from this.
The specific backgrounds that have been treated in more detail are:
\begin{equation}
\label{eq:Processes}
\begin{array}{lllrr}
	\textrm{{\bf Initial}} &&\textrm{{\bf Final} }&\textrm{{\bf Actual \# of events}}&\textrm{{\bf Simulated \# of events}}\\
	f \bar f	&\rightarrow&	\gamma/Z^0	&46200&		59910\\		
	f \bar f	&\rightarrow&	g + \gamma	&202020&	400000\\	
	f \bar f	&\rightarrow&	g + Z^0		&303300&	400000\\	
	f \bar f	&\rightarrow&	\gamma + \gamma	&2276&		10000\\		
	f \bar f	&\rightarrow&	\gamma + Z^0	&5193&		10000\\		
	f \bar f	&\rightarrow&	Z^0 + Z^0	&6732&		34410\\		
	fg      &\rightarrow&   f + \gamma		&886500&	400000\\	
	fg      &\rightarrow&   f + Z^0 		&1261500&	400000\\	
	Z^0Z^0  &\rightarrow&   Z^0 + Z^0		&0&		0     \\	
	WW      &\rightarrow&   Z^0 + Z^0       	&3&		0      \\	
	gg      &\rightarrow&   \gamma + \gamma 	&66&		10000     \\	
	gg      &\rightarrow&   g + \gamma      	&57&		10000	\\	
	\gamma+jet&\rightarrow&	\gamma\gamma		&564450&	400000
\end{array}
\end{equation}
Here actual \# of events is the number of events that would be produced for an integrated
luminosity of 300 fb$^{-1}$.
Note that the last line represents the fake photons that come from QCD processes, as
described in \ref{sec:FalsePho}.
As can be seen the $Z^0Z^0  \rightarrow   Z^0 + Z^0$ and $WW      \rightarrow   Z^0 + Z^0$
have a very small cross-section. In fact, they are so tiny for the 
given cuts (see~section \ref{sec:cuts}) that statistically, there will not be
a single such event detected in ATLAS. These events are excluded as they will not contribute
to any of the backgrounds. For most processes, an excess of events has been produced.
The exceptions are $fg      \rightarrow   f + \gamma$ and $fg      \rightarrow   f + Z^0$
and the photon-jet confusion in the QCD background.
However, this mixup of photons and jets is only considered for the $X\rightarrow \gamma\gamma$ channel.
That these processes are underproduced can be motivated by the small contribution
they add to the total background.

For the sake of completeness, all these backgrounds have been treated
with all the selection criteria of the different processes of interest, see section \ref{sec:SelCrit}.
However, only a few of them will actually produce a background for a certain signal. These
are the backgrounds presented in the figures.


\subsubsection{Background for $X \rightarrow \gamma\gamma$ at $m_X\sim 1$ TeV}
\begin{figure}
        \centering
        \includegraphics[width=3in]{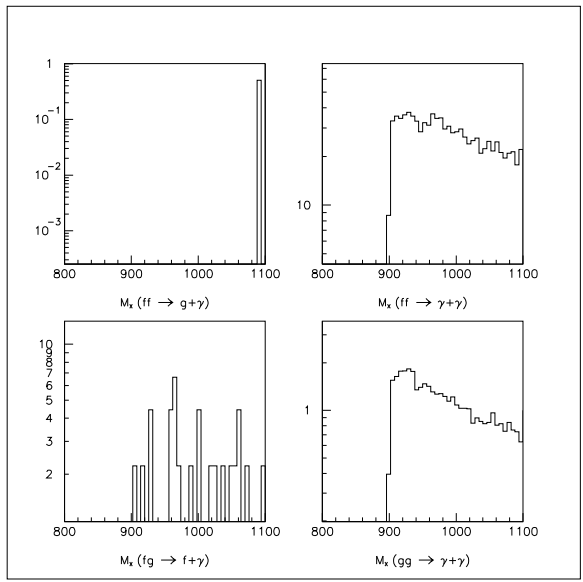}
        \caption{Histograms for the contributing backgrounds for the process $X\rightarrow \gamma\gamma$ for $m_X\sim~1000$ GeV,
		after preselection cuts.}
        \label{fig:BGGamGam}
\end{figure}

The background of this process is presented in figure \ref{fig:BGGamGam}.
The signals are normalized to an integrated luminosity of 300 fb$^{-1}$.
We notice that the only processes that actually give a background are
$\bar ff      \rightarrow   \gamma + \gamma$
and
$gg      \rightarrow   \gamma + \gamma$.
Even though their cross-sections are not so high, they have a final state
that is irreducible. Hence, these are the two processes that we will keep when
we calculate the backgrounds for different axion masses. Of these two,
$\bar ff      \rightarrow   \gamma + \gamma$
is largest, which is only to expect in proton-proton collisions.

In addition to the direct backgrounds we have also simulated the QCD background
to account for the possibility of confusion of a jet and a photon, as discussed in
chapter \ref{sec:FalsePho}. This is simulated just like all the other processes except
that the cross-section is taken to be 3000 times\footnote{
	This is motivated by the result, which does not contain much background 
	(figure~\ref{fig:BGGamGam}).
} less (otherwise the number of 
events would be enormous). When the selection criteria are applied, the invariant
mass is calculated from the photon and jet with highest transverse momenta.

\subsubsection{Background for $X \rightarrow \gamma Z^0 \rightarrow \gamma \bar ll$ at $m_X\sim 1$ TeV}
\begin{figure}
        \centering
        \includegraphics[width=3in]{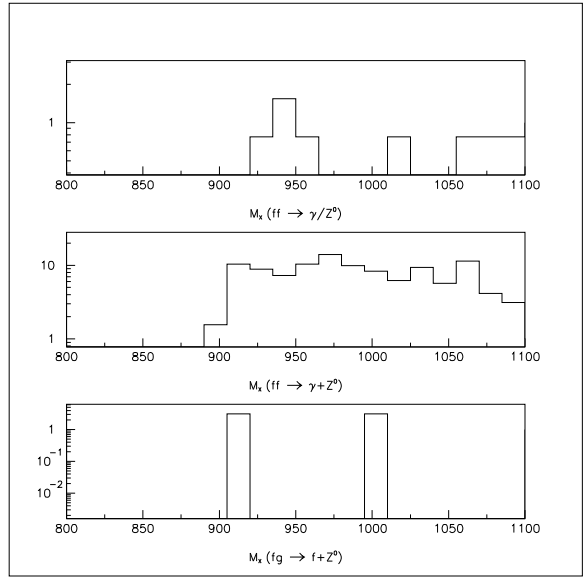}
        \caption{Histograms for the contributing backgrounds for the process
		$X\rightarrow Z^0\gamma\rightarrow l\bar l \gamma$ for $m_X\sim~1000$ GeV,
		after preselection cuts.}
        \label{fig:BGGZll}
\end{figure}

The background of this process is presented in figure \ref{fig:BGGZll}.
The signals are normalized to an integrated luminosity of 300 fb$^{-1}$ and the cuts (see~section \ref{sec:cuts})
have been applied.
The only backgrounds that contribute to the signal are
$\bar ff      \rightarrow   \gamma/Z^0$
and
$\bar ff      \rightarrow   \gamma + Z^0$.
These two signals will be kept for analysis of different axion masses.
We do not consider the possible confusion of a jet with a photon because
the probability of production of a $Z^0$ with the imposed cuts is very low in these
processes.
In analogy with the $X\rightarrow \gamma\gamma$ process, the $\bar ff      \rightarrow   \gamma + Z^0$
gives the irreducible background.

\subsubsection{Background for Other Processes at $m_X\sim 1$ TeV}
As predicted in chapter \ref{ch:CaEx}, the background for axion decay to
jets is tremendous, several order of magnitudes larger than the signals themselves.
The one with lowest background (but still very large) is
$X \rightarrow Z^0Z^0\rightarrow jet+jet+\bar l l$, but
unfortunately the cross-section is too small.
Thus all final states that include jets are worthless for detecting the axion.
The only process left, $X \rightarrow Z^0Z^0\rightarrow \bar l l\bar l l$ 
has a vanishing cross-section and will give about one event in 55 years of LHC
operation at high luminosity (total: 6000 fb$^{-1}$).





\subsection{Significant Background Processes for $m_X\sim 800$ GeV}
The cross-sections are calculated with the two initial cuts, $p_T>300$ GeV
and $\sqrt{\hat s}>700$ GeV (see~section \ref{sec:cuts}). 
The signals are normalized to an integrated luminosity of 300 fb$^{-1}$.


\subsubsection{Background for $X \rightarrow \gamma\gamma$ at $m_X\sim 800$ GeV}
\begin{figure}
        \centering
        \includegraphics[width=3in]{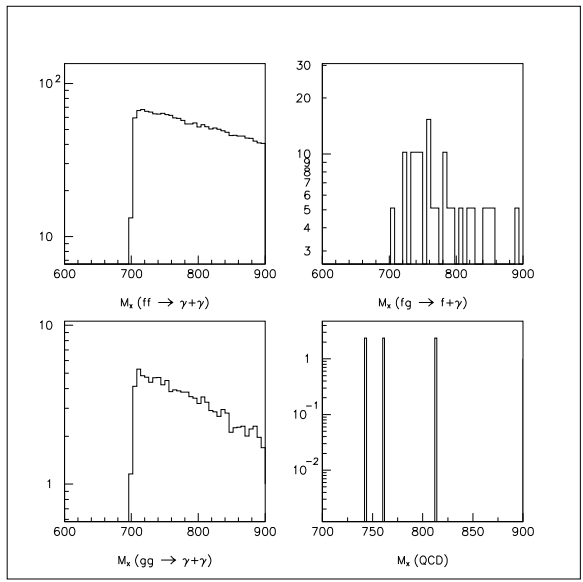}
        \caption{Histograms for contributing backgrounds for $X\rightarrow \gamma\gamma$ at $m_X \sim 800$ GeV,
		after preselection cuts.}
        \label{fig:800GG}
\end{figure}

As we found in section \ref{sec:mX1000} there are only two important processes
for $\gamma\gamma$ production. However, we have also considered the QCD background
as this is known to increase dramatically for lower cuts.
The result is presented in figure \ref{fig:800GG}.  Except for the lower
cutoff in $\sqrt{\hat s}$, it is the same background 
as for the case of $m_X\sim 1000$ GeV.

\subsubsection{Background for $X \rightarrow \gamma Z^0 \rightarrow \gamma \bar ll$ at $m_X\sim 800$ GeV}
\begin{figure}
        \centering
        \includegraphics[width=3in]{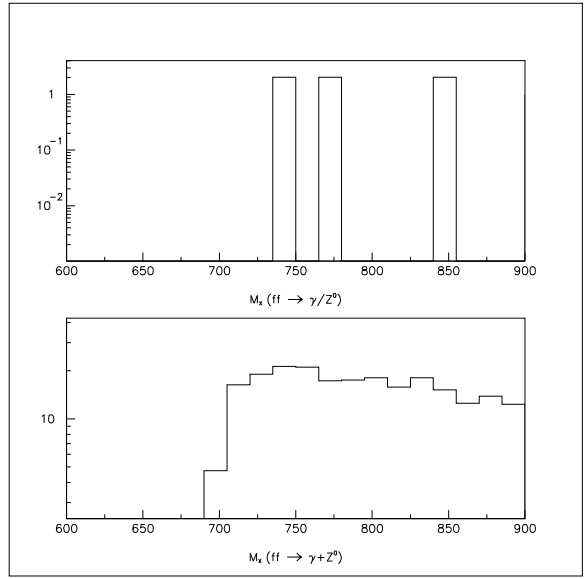}
        \caption{Histograms for contributing backgrounds for $X\rightarrow Z^0\gamma\rightarrow\bar l l\gamma$ at $m_X \sim 800$ GeV,
		after preselection cuts.}
        \label{fig:800ll}
\end{figure}

As we found in section \ref{sec:mX1000} there are only two important processes
for $\gamma\bar ll$ production. 
The result is presented in figure \ref{fig:800ll}. No spectacular deviations from
the case $m_X\sim 1000$ GeV is noted.

\subsubsection{Background for Other Processes at $m_X\sim 800$ GeV}
For completeness, the other possible final states for axion decay should
be considered also for this cut. Due to the weak signal of $Z^0Z^0$
production and the enormous background for
$X\rightarrow \gamma Z^0 \rightarrow \gamma +jet+jet$ the above mentioned
processes are the only ones that can be used for detection and identification of the
hypercharge axion. 
If the integrated luminosity and the energy could be drastically increased they could be interesting,
but in LHC they would hardly show up at all, as discussed in section \ref{sec:OthProc}.

\section{Detection of the Hypercharge Axion in \mbox{ATLAS}}
\label{sec:BgSig}
In this section we will study the two main decay channels of the axion,
$X \rightarrow \gamma\gamma$ and {$X \rightarrow \gamma Z^0 \rightarrow \gamma ll$.
We will study the signals with the cuts and apply the backgrounds to see whether
the axion can be expected to be discovered. The condition for discovery will be that
the signal should contain more events than five times the standard deviation
of the background (i.e., five times the square-root of the number of background events in
the mass-window), and that the number of events from the signal is $> 10$:
\begin{equation}
	N_{ev} > 5\times \sqrt{B}\quad	\textrm{and}	\quad	N_{ev} > 10.
\end{equation}
The ratio $N_{ev}/\sqrt{B}$ is called significance.
The number of events is integrated between $m_X-20$ GeV and $m_X+20$ GeV.
Note that the signal is expressed in units of $(\frac{\textrm{TeV}}{M_Y})^2$,
which means that the signal will go down rapidly with increasing $M_Y$.
Note that often in the theoretical models (see~chapter \ref{ch:BaMo}) 
$M_Y \gg M_X$.

Two different masses of the axion are discussed, 800 GeV and 1000 GeV.

\subsection{Signal and Background for $X \rightarrow \gamma\gamma$}
For both masses, the signal is rather clean, and pretty strong.

For $m_X=1$ TeV, the signal and the backgrounds are presented in figure \ref{fig:1000GG+bg} and
we can clearly see the signal over the background. 
In the window $980<m_X<1020$ GeV we have about 204 background events and 210 signal events.
This gives a significance of 14.7 sigma with $M_Y=1$ TeV (see beginning of section).
\begin{figure}
        \centering
        \includegraphics[width=3in]{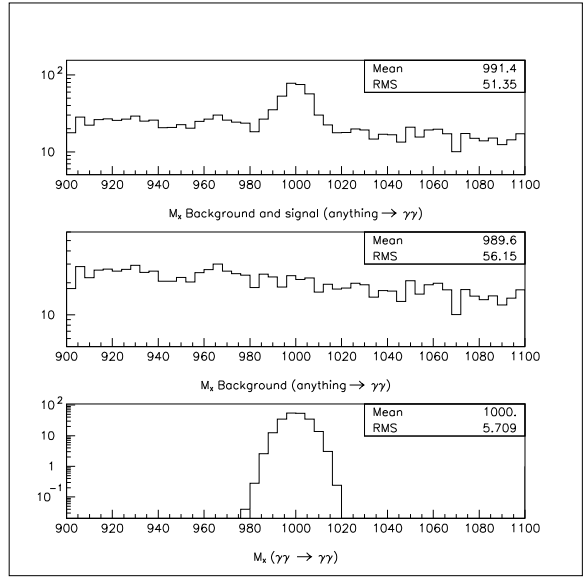}
        \caption{Histograms for signal and background for $X\rightarrow\gamma\gamma$ at $M_Y=1000$ GeV, $m_X = 1000$ GeV,
		after preselection cuts.}
        \label{fig:1000GG+bg}
\end{figure}

For $m_X = 800$ GeV the number of events is higher but so is the background,
as can be seen in figure \ref{fig:800GG+bg}. 
In the window $780<m_X<820$ GeV we have about 405 background events and 255 signal events.
The significance is about 12.7 sigma, which means that the signal is possible to detect with $M_Y=1$ TeV
(see beginning of section).
We remark that the background grows faster than the signal as the mass goes down.
\begin{figure}
        \centering
        \includegraphics[width=3in]{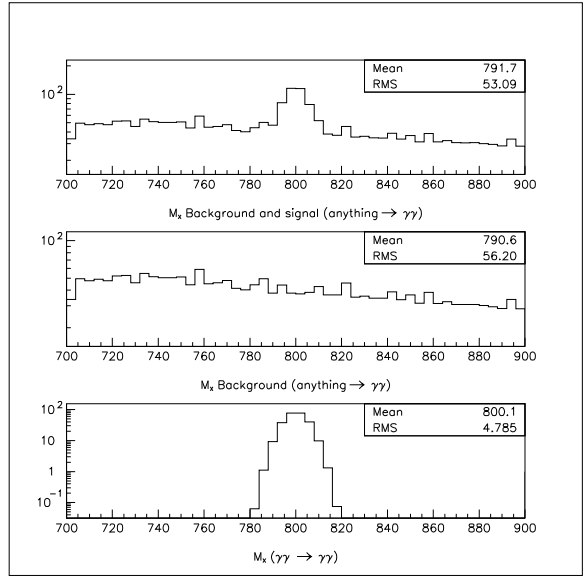}
        \caption{Histograms for signal and background for $X\rightarrow\gamma\gamma$ at $M_Y=1000$ GeV, $m_X = 800$ GeV,
		after preselection cuts.}
        \label{fig:800GG+bg}
\end{figure}

\subsection{Signal and Background for $X \rightarrow \gamma Z^0 \rightarrow \gamma \bar ll$}
This signal is also clean.
The only drawback is that the branching ratio for $Z^0\rightarrow \bar l l$
is so low.

The background is very small for $m_X=1$ TeV but so is the signal as
can be seen in figure \ref{fig:1000ll+bg}.
In the window $980<m_X<1020$ GeV we have about 26 background events and 11.4 signal events.
The signal is hardly
above the background and we have a significance of 2.2 sigma for $M_Y=1$ TeV
(see beginning of section).

\begin{figure}
        \centering
        \includegraphics[width=3in]{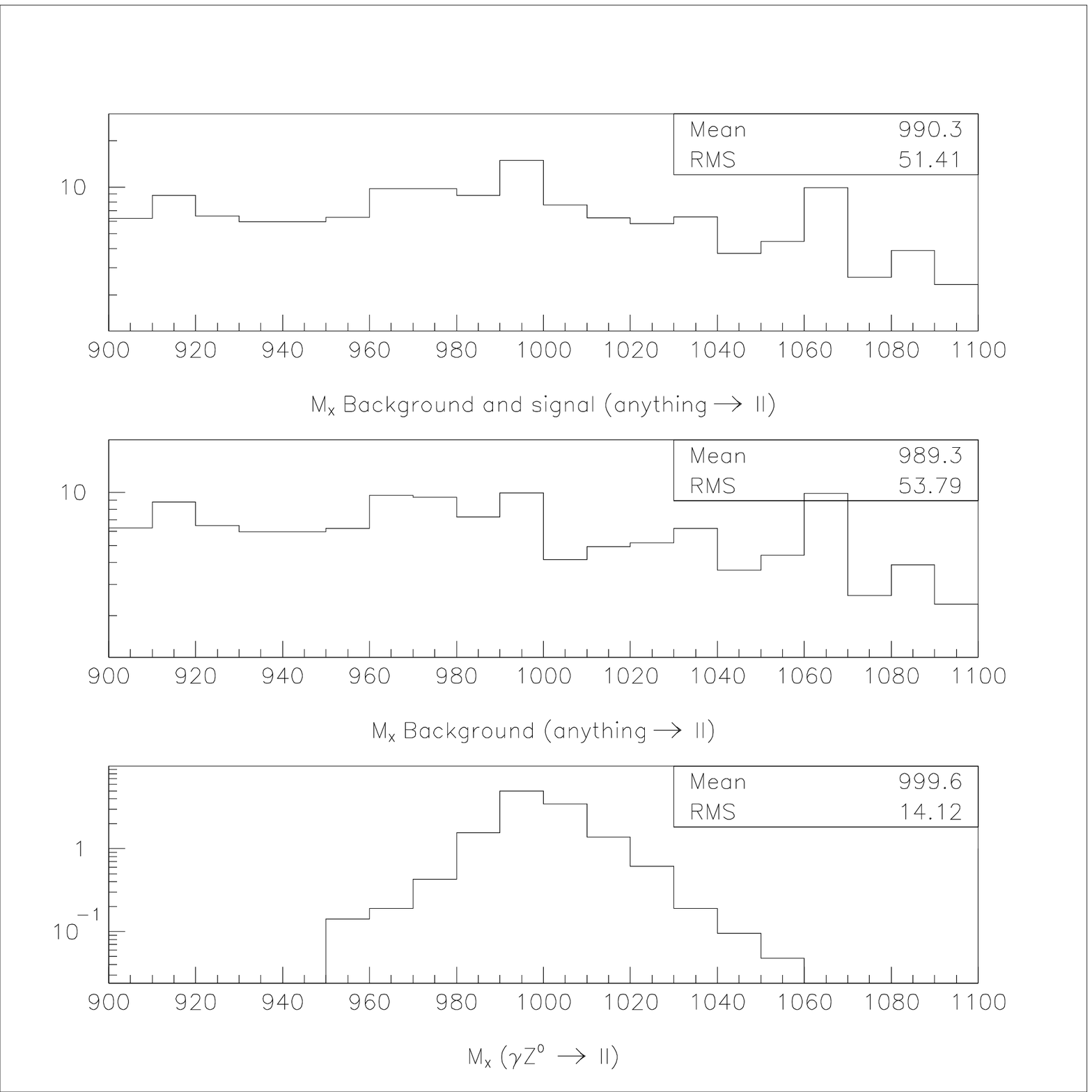}
        \caption{Histograms for signal and background for $X\rightarrow Z^0\gamma\rightarrow\bar l l\gamma$ at $M_Y=1000$ GeV, $m_X = 1000$ GeV,
		after preselection cuts.}
        \label{fig:1000ll+bg}
\end{figure}

For a lower axion mass, $m_X=800$ GeV, the signal is not clearly visible either (see figure \ref{fig:800ll+bg}).
In the window $780<m_X<820$ GeV we have about 45 background events and 15 signal events.
The significance is 2.2 sigma for $M_Y=1$ TeV (see beginning of section).
\begin{figure}
        \centering
        \includegraphics[width=3in]{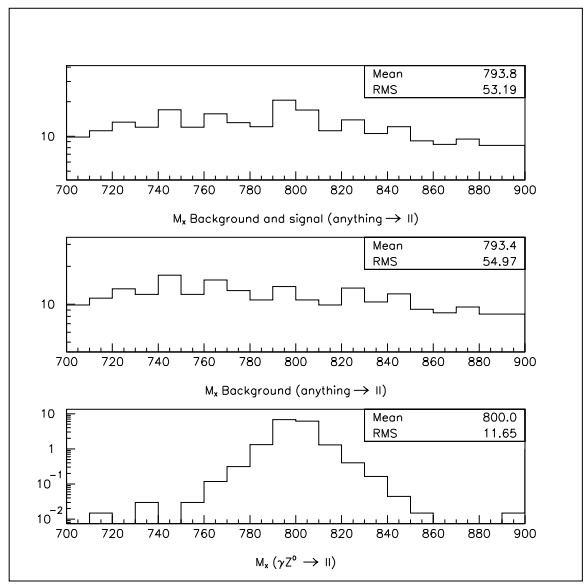}
        \caption{Histograms for signal and background for $X\rightarrow Z^0\gamma\rightarrow\bar l l\gamma$ at $M_Y=1000$ GeV, $m_X = 800$ GeV,
		after preselection cuts.}
        \label{fig:800ll+bg}
\end{figure}


\section{Discussion and Conclusions}
In the previous sections we have seen that there are basically two processes
that can be used for detection of the hypercharge axion; $X\rightarrow \gamma\gamma$
and $X\rightarrow Z^0\gamma\rightarrow\bar ll \gamma$. The former signal
has the higher signal-to-background ratio. In the case of $m_X=800$ GeV and $M_Y=1000$ GeV
the axion would be observed (a 14.7 sigma significance).
The other signal, $X\rightarrow Z^0\gamma\rightarrow\bar ll \gamma$, is more difficult
to distinguish and for the same parameters we have a significance 
of 2.2 sigma. This is not enough for discovery, in the case of a clear signal for the $X\rightarrow \gamma\gamma$, 
but it could support a discovery and verify the branching ratios.

If we want to be certain that it is really the hypercharge axion that is
detected, the branching ratios of $X\rightarrow \gamma\gamma$ and
$X\rightarrow Z^0\gamma\rightarrow\bar ll \gamma$ have to be compared.
The ratio $\Gamma_{\gamma\gamma}/\Gamma_{\bar ll\gamma}$ should be
28 for the hypercharge axion. The cross-section in itself cannot
serve as a certain identification, though well as an indication,
because the cross-section is proportional to $1/M_Y^2$ and $M_Y$
is an unknown parameter with dimension of mass.
If all three decay channels are discovered, then it is a strong indication
that we have discovered a pseudoscalar with the right coupling
to the hypercharge. At that stage, the mass-scale $M_Y$ can be used
to determine theoretically whether the pseudoscalar could amplify the hypermagnetic
fields or not \cite{BruOakExt}.

The scenario of $m_X=800$ GeV and $M_Y=1000$ GeV is taken from the
article by Brustein and Oaknin \cite{BruOakColl}, which gave the original
idea of this thesis. The cross-sections in both their and our case are
about 10 fb$^{-1}$.

There is also a theoretical argument why a particle in the mass
region $300<m_X<1000$ GeV should be the hypercharge axion. It is not
expected that there exist any particles in this mass range with electroweak
couplings. The Higgs could in principle have this mass but this is
outruled by the LEP (Large Electron Positron collider) high
precision electroweak measurement. Furthermore,
for a Higgs of this mass, the decay channel to two photons would 
be non-existent anyway. The Higgs would also have other, stronger, decay 
channels (like $W^+W^-$).

Note that in the calculations in this chapter, $M_Y$ is supposed to
be 1 TeV, or, in other words, everything is expressed in terms
of $(\frac{\textrm{TeV}}{M_Y})^2$. The parameter $M_Y$ is
an energy scale, which is supposed to be much larger than $m_X$
but in this thesis we disregard this to discover the limits
of hypercharge axion detection in ATLAS. For $M_Y\lesssim 1.6$ TeV
detection should be possible but hardly much above that.
The precise limit depends on the mass of the axion too.

This thesis covers the possibility of discovering the hypercharge
axion in ATLAS, but there are several points that could be elaborated
and treated in more detail. Examples are the Weizs\"acker-Williams approximation,
the parton distribution functions and the amount of background
produced. A study of processes and backgrounds could also be made
for other values of the hypercharge axion mass.

There are also several questions in the theory behind the
hypercharge axion that are still obscure. For example
where the original hypermagnetic field came from and
how enough $CP$ violation can be obtained.
	\appendix

\chapter{Glossary}
\label{app:Gloss}
\begin{description}
\item[Anomaly]
	An anomaly is the failure of a symmetry to survive renormalization.
\item[Antimatter]	Antimatter is constituted of \gloss{antiparticles}.
\item[Antiparticle]
	An antiparticle is defined
	as having the opposite \gloss{quantum numbers} as the corresponding particle,
	but the same mass.
	Particles with their quantum numbers zero, except for spin which should be integer, (like the photon) are their own antiparticles.
	Experimentally there is no evidence for the effect of gravity on antimatter,
	but according to Einstein's principle of equivalence there should be no difference to matter.
	However, there are other possible theories, cousins of Einsteins theory of
	gravitation, that predict weaker gravity, or even inverse gravity for
	antiparticles. For a review of antiparticles, see, e.g.,
	J. Eades and F. J. Hartmann, Rev. Mod. Phys. 1999 {\bf 71}, 373-420.
\item[Axial current]
	The axial current is defined as 
	\begin{equation}
		A_\mu(x) = \bar\psi(x)\gamma_\mu\gamma_5\psi(x),
	\end{equation}
	where $\psi(x)$ is the field at space-time point $x$, $\epsilon(x)$ is an
	arbitrary function of $x$ and $\gamma_\alpha,\; (\alpha=1,2,3,4,5)$ are
	defined in appendix \ref{App:Symb}.
\item[Baryogenesis]
	The generation of baryon asymmetry.
\item[Branching ratio]
	The relative probability that a particle will decay in a specific way.
	Example: $Z^0 \rightarrow e^+e^-$ has a branching ratio of about 3 percent.
	The sum of all branching ratios should be unity.
\item[Charge ($C$)]
	$C$ is an operation that reverses the charge of a particle.
\item[Chern-Simons number]
	A \gloss{winding number} for the electroweak Lagrangian.
	The change in Chern-Simons number is proportional to the change
	in baryon number, (see~section \ref{sec:Pendulum}).
\item[Decay channels]	The possible decays of a particle.\\
	Example: $$Z^0 \rightarrow
		u{\bar u},
		d{\bar d},
		s{\bar s},
		c{\bar c},
		b{\bar b},
		t{\bar t},
		e^-e^+,
		\mu^-\mu^+,
		\tau^-\tau^+,
		\nu_e{\bar \nu_e},
		\nu_\mu{\bar \nu_\mu},
		\nu_\tau{\bar \nu_\tau},$$
	where the different processes have different \gloss{branching ratios}.
\item[Electroweak phase transition]
	The electroweak phase transition is when the electroweak symmetry is broken
	and the Higgs particle obtains a vacuum expectation value.
	It is estimated to have occurred about $10^{-10}$ s after big bang, when the temperature
	of the universe was $\sim 100$ GeV.
\item[Final state radiation]
	Final state radiation is decays of particles or emission of gluons or photons that 
	occur after the event. In PYTHIA, this is the process that allow particle
	showers to evolve.
\item[Goldstone boson]
	A zero spin, zero mass particle associated with symmetry breaking.
\item[Higgs boson]
	A hypothetical particle required in the standard model of particle physics.
	The Higgs boson explains why $W^\pm$, and $Z^0$ have a mass.
\item[Initial state radiation]
        Initial state radiation is decays of particles or emission of gluons or photons that
        occur before the event.
\item[Massshell]
	On the mass shell means that the particle is real ($M^2=E^2-|\vec p|^2$).
	Off the mass shell means that it is virtual ($M^2\ne E^2-|\vec p|^2$).
\item[Parity ($P$)]
	Parity is an operation that gives a particle the same
	properties as if is was observed in a ``point-like mirror''. In other words, the spin
	of the particle will be inversed. $P\left| parity=p\right> = \left| parity=-p\right>$.
	The left-parity operator is represented by $P_L = \frac 12(1-\gamma_5)$,
	where $\gamma_5$ is defined in appendix \ref{App:Symb}. The right-parity
	operator is defined as $P_R = \frac 12(1+\gamma_5)$.
\item[Sphaleron]
	An unstable solution of a partial differential equation.
	A sphaleron process is the process of passing a spahaleron, (see~\ref{sec:Pendulum}.)
\item[Symmetry]
	A symmetry operation does not change the physical solution. 
	For example, the position of the origin in the coordinate system
	does not change the physical solution to problem.
\item[Winding number]
	A winding number is a characteristic that does not change during
	continuous transformations, if a certain point is avoided.
	A trivial example of this is found in complex analysis:
	\begin{equation}
		I(\gamma_{AB},p) = \frac{1}{2\pi i}\int_{\gamma_{AB}} \frac{dz}{z-p}
	\end{equation}
	where the value of the integral is independent of the curve $\gamma_{AB}$
	as long as the points $A$ and $B$ are the same and $p$ is avoided. 

	The winding number is a sort of shortcut which lets us know the
	consequences of micro-level rules without elaborate calculations.

\item[Quantum numbers]	The numbers that can be said to best describe the state of
	a particle. Examples: electric charge ($Q$), lepton number ($L$),
	baryon number ($B$), parity ($P$), spin ($S$), isospin ($I$),
	strangeness ($S$), and charge conjugation ($C$).
\end{description}
\chapter{List of Symbols}
\label{App:Symb}
\section{In Equations}
\begin{tabular}{ll}
	$X$	&	Pseudoscalar axion\\
	$m_X$	&	Mass of the axion\\
	$M_Y$	&	Mass scale\\
	$Y_{\mu\nu}$&	Hypercharge field strength (from $U(1)$)\\
	$\tilde Y_{\mu\nu}$&	The dual of $Y_{\mu\nu}$, $\tilde Y_{\mu\nu} = \epsilon_{\alpha\beta\mu\nu}Y^{\alpha\beta}$\\
	$\epsilon^{\mu\nu\ldots}$&	The permutation symbol,
					defined as +1 for even permutationes of $\epsilon^{012\ldots}$\\
		&			-1 for odd permutationes, and 0 for other values of the indices\\
	$A_\mu$	&	Hyperphoton vector\\
	$Z_\mu$	&	$Z^0$ vector\\
	$m_Z$	&	Mass of $Z^0$\\
	$H$	&	Hamiltonian\\
	$T$	&	Temperature\\
	$E$	&	Energy\\
	$B=n_B$	&	$n_B = n_b - n_{\bar b}$, baryon number:\\
		&	difference between the number of baryons and antibaryons\\
	$L$	&	Lepton number\\
	$s$	&	Entropy\\
	$\Gamma$&	Transition rate or branching ratio\\
	$N_{CS}$&	Chern-Simons (winding) number\\
	$S_E$	&	Euclidian action\\
	$H_Y$	&	Hypermagnetic field\\
	$E_Y$	&	Hyperelectric field\\
	$J_B$	&	Baryon current\\
	$\mu$	&	Chemical potential\\
\end{tabular}
\begin{tabular}{ll}
	$\gamma$&	Photon\\
	$l$	&	Lepton\\
	$g$	&	Gluon\\
	$\gamma$&	Photon\\
	$f$	&	Fermion (quark or lepton)\\
	$q$	&	Quark \\
	$jet$	&	Jet of particles, produced by a quark or a gluon\\
	$t$	&	Top quark\\
	$b$	&	Bottom quark\\
	$p$	&	Momentum\\
	$p_T$	&	Transverse momentum, $p_T \equiv \sqrt{p_x^2+p_y^2}$ if $z$ is the beam direction\\
	$\cal L$&	Luminosity\\
	$\sigma$&	Cross-section\\
	$t$	&	Time\\
	$n_f$	&	Number of families of quarks/leptons, believed to be three\\
	$i$	&	Imaginary number, defined as $i^2=-1$ \\
	$\gamma$ matrices&	
			$\gamma_0 = \left[\begin{smallmatrix} 1 & 0 & 0 & 0 \\ 0 & 1 & 0 & 0 \\ 0 & 0 & -1 & 0 \\ 0 & 0 & 0 & -1 \end{smallmatrix}\right]$
			$\gamma_5 = \left[\begin{smallmatrix} 0 & 0 & 1 & 0 \\ 0 & 0 & 0 & 1 \\ 1 & 0 & 0 & 0 \\ 0 & 1 & 0 & 0 \end{smallmatrix}\right]$\\
		&	$\gamma_1 = \left[\begin{smallmatrix} 0 & 0 & 0 & 1 \\ 0 & 0 & 1 & 0 \\ 0 & -1 & 0 & 0 \\ -1 & 0 & 0 & 0 \end{smallmatrix}\right]$
			$\gamma_2 = \left[\begin{smallmatrix} 0 & 0 & 0 & -i \\ 0 & 0 & i & 0 \\ 0 & i & 0 & 0 \\ -i & 0 & 0 & 0 \end{smallmatrix}\right]$
			$\gamma_3 = \left[\begin{smallmatrix} 0 & 0 & 1 & 0 \\ 0 & 0 & 0 & -1 \\ -1 & 0 & 0 & 0 \\ 0 & 1 & 0 & 0 \end{smallmatrix}\right]$
\end{tabular}

%
%
%

\section{List of Constants}
\begin{tabular}{lll}
        $\theta_W$&	28.7 degrees&	The weak mixing angle $sin \theta_W  \approx 0.23 $\\
	$\hbar$	&	1.054571597$\times 10^{-34}$ Js&	$\hbar=h/2\pi$, where $h$ is Planck's constant\footnotemark\\
	$c$	&	299792458 ms$^{-1}$&	Speed of light in vacuum\footnotemark[\value{footnote}]\\
	$k_B$	&	1.3806503$\times 10^{-23}$ JK$^{-1}$&	Boltzmann's constant\footnotemark[\value{footnote}]\\
	eV	&	1.6022$\times 10^{-19}$ J&		Electron Volt
\end{tabular}
\footnotetext{
	Unless otherwise specified, we set $\hbar=c=k_B=1$}

\section{Abbreviations}
\begin{tabular}{ll}
	ATLAS	&	A Toroidal LHC ApparatuS\\
	ATLFAST	&	A FORTRAN program to approximately simulate the ATLAS-detector\\
	BAU	&	Baryon Asymmetry of the Universe\\
	C	&	Charge-reverse operator\\
	CERN	&	European laboratory for particle physics\\
		&	(earlier: Conseil Europ\'een pour la Recherche Nucl\'eaire)\\
	CP	&	Charge-Parity operator\\ 
	CPT	&	Charge-Parity-Time operator\\
	EW	&	Electro Weak\\
	EWPT	&	Electro Weak Phase Transition\\
	HCA	&	HyperCharge Axion\\
	LHC	&	Large Hadron Collider\\
	LEP	&	Large Electron Positron collider\\
	LTU	&	Lule\aa~University of Technology (in Swedish: Lule\aa~Tekniska Universitet)\\
	MSSM	&	Minimal Supersymmetric Standard Model\\
	P	&	Parity operator\\
	PYTHIA	&	A FORTRAN program to simulate collisions between particles\\
	QCD	&	Quantum ChromoDynamics, the field theory for strong interactions\\
	QED	&	Quantum ElectroDynamics, the field theory for electromagnetic interactions\\
	SUSY	&	SUper SYmmetry\\
	T	&	Time operator\\
\end{tabular}
\pagebreak


\begin{thebibliography}{40}
        \bibitem {Nash} C. Nash, 1978, {\it Relativistic Quantum Fields},
                Academic Press.
	\bibitem {PeskSch} M. E. Peskin, D. V. Schroeder, 1995, {\it An Introduction to Quantum Field Theeory}, Addison Wesley Publishing Company.
        \bibitem {BruOakColl} R. Brustein and D. H. Oaknin, 1999, hep-ph/9906344.
	\bibitem {EWBaGe} M. Trodden, 1999, Rev. Mod. Phys., {\bf 71}, 1463.
	\bibitem {Steigman} G. Steigman, 1976, Ann. Rev. Astron. Astrophys.
		{\bf 14}, 336 (from \cite{EWBaGe}).
	\bibitem {Stecker} F. W. Stecker, 1985, Nucl. Phys. {\bf B252}, 25 (from \cite{EWBaGe}).
	\bibitem {Cohen} A. G. Cohen, A. De~R\'ujula, and S. L. Glashow, 1998,
		Astrophys. J. {\bf 495}, 539.
	\bibitem {Sakharov} A. D. Sakharov, 1967, Zh. Eksp. Teor. Fiz. Pis'ma Red. {\bf 5},
		32 [JETP Lett. {\bf 5}, 24 (1967)] (from \cite{EWBaGe}).
	\bibitem {Kolb} E. W. Kolb and S. Wolfram, 1980, Nucl. Phys. {\bf B172}, 224.
	\bibitem {Kabir} P. T. Kabir, 1965, {\it Symmetries in Elementary Physics}, 
	edited by A.~Zichichi, Academic Press, New York. 
	\bibitem {Eades} J. Eades and F. J. Hartman, 1999, Rev. Mod. Phys. {\bf 71}, 373.
	\bibitem {Klinkhammer} F. R. Klinkhammer and N. S. Manton, 1984, Phys. Rev. {\bf D30}, 2212.
	\bibitem {Moore} G. D. Moore, C. Hu and B. Mueller, 1998, Phys. Rev.  {\bf D58}, 045001.
	\bibitem {QCDaxion} R. D. Peccei and H. R. Quinn, 1977, Phys. Rev. {\bf D16}, 1791\\
			S. Weinberg, 1978, Phys. Rev. Lett. {\bf 40}, 223\\
			M. Dine, 1981, Phys. Lett. {\bf B104}, 199.
	\bibitem {PriMF} M. Giovanni and M. E. Shaposhnikov, 1998, Phys. Rev. Lett. {\bf 80}, 22.
	\bibitem {BruOakAmp} R. Brustein and D. H. Oaknin, 1999, hep-ph/9901242.
	\bibitem {BruOakExt} R. Brustein and D. H. Oaknin, 2000, hep-ph/00009009.
	\bibitem {PapaHiggs} E. Papageorgiu, 1995, Phys. Lett. {\bf B352}, 394.
	\bibitem {PYTHIA} T. Sj\"ostrand, 1994, Computer Physics Commun. {\bf 82}, 74.
	\bibitem {PDG} Particle Data Group, 2000, Eur. Phys. J. {\bf C15}, 1.
	\bibitem {PDF} M. Gluck, E. Reya, and A. Vogt, 1990, Z. Phys. {\bf C48}, 471.
	\bibitem {Higgs} A. Djouadi, 1999, Int. J. Mod. Phys. {\bf A10}, 1.
	\bibitem {TDR} ATLAS Detector \& Physics Performance Technical Design Report, 1999, CERN/LHCC/99-14.
\end{thebibliography}
\end{document}